\documentclass[12pt]{iopart}
\usepackage{graphicx}%
\usepackage{multirow}%
\usepackage{amsmath,amssymb,amsfonts}%
\usepackage{amsthm}%
\usepackage{mathrsfs}%
\usepackage[title]{appendix}%
\usepackage{xcolor}%
\usepackage{textcomp}%
\usepackage{manyfoot}%
\usepackage{booktabs}%
\usepackage{algorithm}%
\usepackage{algorithmicx}%
\usepackage{algpseudocode}%
\usepackage{listings}%
\usepackage{bm}
\usepackage{color}
\usepackage{MnSymbol}
\usepackage{enumerate}
\usepackage{bbold,soul}
\usepackage{lipsum}
\usepackage{array}
\usepackage{hyperref}
\usepackage{textgreek}
\usepackage{tabularx}
\usepackage{graphics,epsfig,subfigure}
\usepackage{cleveref}
\newcommand{\beq}{\begin{equation}}
\newcommand{\eeq}{\end{equation}}
\newcommand{\kd}{\kappa}
\newcommand{\ks}{K}
\newcommand{\gu}{\gamma_\uparrow}
\newcommand{\gd}{\gamma_\downarrow}
\newcommand{\mg}{\bar\gamma}
\newcommand{\bg}{\breve\gamma}
\newcommand{\rtp}{z}
\renewcommand{\ul}[1]{\underline{{#1}}}
\newcommand{\noise}{\Xi}
\newcommand{\bra}[1]{\langle{#1}|}
\newcommand{\ket}[1]{|{#1}\rangle}
\newcommand{\sq}[1]{\left[ {#1} \right]}

\newcommand{\coh}{\mathcal{C}}
\newcommand{\cohc}{\coh^{\rm c}}
\newcommand{\cohnc}{\coh^{\rm nc}}
\newcommand{\ccc}{\mathcal{A}}
\newcommand{\phis}{\Phi}
\newcommand{\phid}{\phi}
\newcommand{\dd}{{\rm d}}

\newcommand{\half}{\frac{1}{2}}
\newcommand{\ddt}{\tau} 
\newcommand{\tsr}{\tau_{\rm sr}}

\newcommand{\coeftsr}{\theta_{\rm sr}}
\newcommand{\meas}{{\mu}}
\newcommand{\elam}{\lambda}

\newcommand{\delthe}{\delta \Theta}
\newcommand{\delkd}{\delta\kappa}
\newcommand{\xd}{X}
\newcommand{\info}{Y}
\newcommand{\hfunc}{\mathbf{H}}
\newcommand{\ffun}{\mathbf{F}}
\newcommand{\newF}{\mathbf{F}}
\newcommand{\newFepsilon}{\mathbf{F}}
\newcommand{\deadt}{\ddt_{\rm dd}}

\newcommand{\measerr}{\epsilon}
\newcommand{\sqdepha}{\chi}
\newcommand{\opt}{^\star}
\newcommand{\tp}{^{\top}}
\newcommand{\ie}{\emph{i.e.}}

\definecolor{nblue}{rgb}{0.06,0.3,0.73}
\definecolor{nblack}{rgb}{0,0,0}
\definecolor{nred}{rgb}{0.9,0.1,0.1}
\definecolor{nmagenta}{rgb}{0.7,0.0,0.3}
\definecolor{neditcolor}{rgb}{0.3,0.3,0.9}

\newcommand{\eg}{\emph{e.g.}}

\theoremstyle{thmstyleone}%
\begin{document}

\title[Imperfection analyses for random-telegraph-noise mitigation using spectator qubits]{Imperfection analyses for random-telegraph-noise mitigation using spectator qubits}

\author{Y. Liu$^{1,2}$, A. Chantasri$^{1,3,*}$, H. Song$^{1,4}$ and H. M. Wiseman$^{1,*}$}
\address{$^{1}$Centre for Quantum Dynamics, Griffith University, Kessels Rd, Brisbane, 4111 Queensland, Australia}
\address{$^2$School of Engineering, University of Newcastle, University Dr, Newcastle, 2308 New South Wales, Australia}
\address{$^3$Optical and Quantum Physics Laboratory, Department of Physics, Faculty of Science, Mahidol University, Rama VI Rd, Bangkok, 10400, Thailand}
\address{$^{4}$Qian Xuesen Laboratory of Space Technology, China Academy of Space Technology, Youyi Rd, Beijing, 100094, China}
\address{$^{*}$Authors to whom any correspondence should be addressed}
\eads{\mailto{yaananliu@gmail.com}, \mailto{a.chantasri@griffith.edu.au}, \mailto{shtfc@163.com}, and \mailto{h.wiseman@griffith.edu.au}}
\vspace{10pt}
\begin{abstract}
Spectator qubits (SQs) for random-telegraph noise mitigation have been proposed by Song \emph{et al.}, Phys.~Rev.~A, {\bf 107}, L030601 (2023), where an SQ operates as a noise probe to estimate optimal noise-correction control on the hard-to-access data qubits. It was shown that a protocol with adaptive measurement on the SQs and a Bayesian estimation-based control can suppress the data qubits' decoherence rate by a large factor with quadratic scaling in the SQ sensitivity. However, the protocol's practicality in real-world scenarios remained in question, due to various sources of imperfection that could affect the performance. We therefore analyze here the proposed adaptive protocol under non-ideal conditions, including parameter uncertainties in the system, efficiency and time delay in readout and reset processes of the SQs, and additional decoherence on the SQs. We also explore analytical methods of Bayesian estimation in the time domain and generalize the map-based formalism to non-ideal scenarios. This allows us to derive imperfection bounds at which the decoherence suppression remains approximately the same as under ideal conditions.
\end{abstract}

%
\vspace{2pc}


\maketitle

\section{Introduction}\label{sec1}

Decoherence in quantum systems is one of the main challenges to harnessing the power of quantum mechanics for practical applications. For example, qubits rely on maintaining quantum coherence to perform complex calculations efficiently in quantum computers. Many noise mitigation technologies that can prolong the coherence time of computational qubits have been proposed in recent years, 
including the utilisation of spectator qubits (SQs) as probes, to correct noise effects while minimizing direct contact to the data qubits (DQs)~\cite{gupta2020adaptive,majumder2020real,song2023optimized,tonekaboni2023greedy,singh2023mid,youssry2023noise}. In this approach, the DQs used for computing are well-isolated from their environment, and the SQs are assumed to be much more sensitive than the DQs to the target noise, and are also easily measurable. The information about the noise can thus be extracted from the measured results, and the DQs can be corrected based on this information. 
Recently, the idea of SQs has also been extended to a spectator mode~\cite{lingenfelter2023surpassing} dealing with spatially correlated noise. Also, phase error correction by using SQs has been demonstrated in experiments~\cite{singh2023mid}, where an array of Cesium atoms working as spectator qubits correct the phase errors on an array of Rubidium data qubits. In this experiment, it was shown that the SQs can be read out without affecting the DQs by using two different atomic species.

In our previous work~\cite{song2023optimized, tonekaboni2023greedy}, we explored the ultimate limits to the SQ technique. To this end, we considered dephasing in a DQ caused by a random telegraph process (RTP) noise with transition rates $\gu$ and $\gd$ \cite{ItaTok2003, GalAlt2006, culcer2009dephasing, BerGal2009}. By assuming that the SQ is affected by the RTP in the same way as the DQ, but with much greater noise sensitivity, $\ks\gg \kd$ (where $\ks$ and $\kd$ are noise sensitivities of the SQ and the DQ, respectively), we developed the Map-based Optimized Adaptive Algorithm for the Asymptotic Regime (MOAAAR). This algorithm specifies how frequently one should measure the SQ, what basis to measure in at those times, and what correction to implement on the DQ to globally optimize its decoherence rate. In the asymptotic limit, it was shown that the algorithm could reduce the decoherence rate by a multiplicative factor of $1.254(\mg/\ks)^2$, where $\mg=(\gu+\gd)/2$ is assumed $\ll K$. This shows that the SQ can, like dynamical decoupling~\cite{viola1999dynamical,viola2003robust,biercuk2011dynamical,ng2011combining,souza2011robust,medford2012scaling,paz2013optimally,zhang2014protected} and quantum error correction~\cite{Shor1995,Steane1996,Terhal2015}, work arbitrarily well at least for the RTP noise mitigation, within a suitable regime. 

However, all the analyses in Ref.~\cite{tonekaboni2023greedy} and its companion paper~\cite{song2023optimized} were based on the perfect model, where parameters of the system are perfectly known without any uncertainty. In practice, these parameters need to be estimated (\eg, via quantum noise spectroscopy protocols~ \cite{paz2017multiqubit,von2020two,chalermpusitarak2021frame}) and will not be perfectly known. Moreover, they may drift over time. In addition, there may be other imperfections in the system and apparatus, such as non-instantaneous readout, projective measurement errors, and additional decoherence of the SQ, all potentially important effects with spin qubits~\cite{culcer2009dephasing,morello2010single,hanson2007spins}. Therefore, it is useful to analyse how such imperfections or uncertainties in the parameters would affect the proposed algorithm's performance. In this work, we aim to analyse the imperfections and obtain bounds on how large they can be without increasing the DQ decoherence rate by a substantial factor above the (conjectured) ultimate limit of the ideal case~\cite{song2023optimized, tonekaboni2023greedy}.

In our analyses, we use a generalized version of the Bayesian map-based method, relaxing some of the assumptions that were used in the ideal case~\cite{tonekaboni2023greedy,song2023optimized}. This generalization includes time periods during which the SQ cannot be measured and decoherence for the SQ, and allows for parameters that deviate from their nominal values used in the ideal protocol, MOAAAR.  
Moreover, we also introduce a new analytical calculation technique based on the concept of ``Before-time'', $t_{\rm B}$, which is the time interval between the actual flip of the RTP and when that flip is detected. The calculation of DQ's decoherence using the Before-time provides a more elementary and more general approach to re-deriving the decoherence rate and offers deeper insights into the optimality condition derived in the previous work~\cite{tonekaboni2023greedy,song2023optimized}. 

To summarize our results, we can classify them into three classes. First, when there are unknown deviations in the SQ's measurement angle $\delthe$ or the DQ's sensitivity $\delkd$, as long as these are small, the original MOAAAR protocol works well. 
Specifically, we analytically derive (and numerically confirm) tight upper bounds on the size of $\delthe$ and $\delkd$ such that the DQ's decoherence rate is only a constant multiple higher than that achieved by MOAAAR in the ideal case. Second, when the time required to reset the SQ after each measurement, $\tsr$, or the error probability for the SQ measurement, $\epsilon$, are nonzero, we can still apply the MOAAAR protocol for the SQ measurements, but these known imperfections should be taken into account when controlling the DQ. Again, we derive upper bounds on these imperfections if they are not to cause substantial deviation from the ideal DQ decoherence rate, analytically for $\tsr$ and numerically for $\epsilon$. Finally, for a large enough dead time in the DQ measurement apparatus, $\deadt$, the performance of MOAAAR degrades rapidly as $\deadt$ increases. Therefore, we modify the MOAAAR algorithm to take into account $\deadt$, in order to study its effect.

The structure of this paper is as follows. In~\Cref{sec:math}, we introduce the mathematical formulation, including the generalization of the Bayesian map-based method and the concept of Before-time for the adaptive measurement strategy. ~\Cref{sec:imperfections} elaborates all types of imperfections considered in this work. We begin with the analysis of uncertainty in measurement angles in~\Cref{sec:delthe} and proceed to the analysis of uncertainty in  DQ's noise sensitivity in~\Cref{sec:delkappa}. We then investigate imperfections related to  measurement times in~\Cref{sec:meastime} and then analyze measurement errors (or, equivalently, additional decoherence on the SQ) in~\Cref{sec:measerr}. The paper concludes with a recapitulation of our findings and remarks on future work in~\Cref{sec:conclusion}.

\section{Mathematical formulation}\label{sec:math}

In \Cref{sec-dqsq}, we explain the mathematical model of the noise mitigation protocol using a SQ following the previous work \cite{song2023optimized,tonekaboni2023greedy}. Then in Section~\ref{sec-dqcoh}, we extend the calculation of the DQ's coherence to a generalized map-based formalism that is applied to problems of the RTP noise mitigation with arbitrary SQ measurement settings in subsection~\ref{sec-genmap}. 
We present the previously derived Map-based optimized adaptive algorithm in the asymptotic regime (MOAAAR) in subsection~\ref{sec:prederiveH}. In the last subsection, \ref{sec:rederiveH}, we show how the DQ's coherence for the adaptive measurement strategy can be analytically obtained without maps, but using the probability distribution of time between an actual RTP flip and a detected non-null result. Both techniques will be useful for analysing imperfections, in later sections.

\subsection{Data and spectator qubits}\label{sec-dqsq}
We follow the model for noise mitigation in Refs.~\cite{song2023optimized,tonekaboni2023greedy}, consisting of one DQ and one SQ. We assume that the phases of the DQ and the SQ are affected by an unknown environmental noise denoted by $z(t)$ and that the SQ can be measured at various times to obtain information about the noise, which is at a later time, $T$, applying a control pulse (for noise mitigation) on the DQ. The Hamiltonian of this model is
\begin{equation}\label{eq:totalHamiltonian}
    \hat{H}_{\rm tot} = \frac{\kd}{2} \hat{\sigma}_z^{\rm d} \, z(t) + \frac{\ks}{2} \hat{\sigma}_z^{\rm s}\, z(t) + \hat H_{\rm ctrl}^{\rm d}\delta(t-T).
\end{equation}
Here $\hat\sigma_z^{\rm d}$ and $\kappa$ are the $z$-Pauli matrix and noise sensitivity of the data qubit and $\hat\sigma_z^{\rm s}$ and $K$ are those for the spectator qubit. The last term, $\hat H_{\rm ctrl}^{\rm d}\delta(t-T)$ is the control Hamiltonian, which is only implemented on the DQ at the final time $T$. Given the Hamiltonian in Eq.~\eqref{eq:totalHamiltonian}, we can study the noise mitigation in full generality \cite{tonekaboni2023greedy}
by considering the dynamics of the phases for both the DQ and the SQ when they are in equatorial states, defined as 
\begin{subequations}
\begin{align}
    \ket{\phi}^{\rm d} &:= \frac{1}{\sqrt{2}} \left(\ket{+1}_z^{\rm d} + e^{i \phi} \ket{-1}_z^{\rm d} \right),\\
    \ket{\phis}^{\rm s} &:= \frac{1}{\sqrt{2}} \left(\ket{+1}_z^{\rm s} + e^{i \phis} \ket{-1}_z^{\rm s} \right).
\end{align}
\end{subequations}
Here $\ket{\pm 1}_z^{\rm s,d}$ are eigenstates of the $z$-Pauli matrices of the two qubits. Without loss of generality, we can set the DQ's phase at the initial time as a zero-phase state, $\ket {\phid=0}^{\rm d}$, which, under the influence of the noise, will evolve to a state 
\beq\label{eq:dataphase}
\ket {\phid(X)}^{\rm d}=\exp\left(- i \tfrac{\kappa}{2}\hat \sigma_z^{\rm d} \! \int_0^t \! \dd s\,  z(s) \right)\ket {\phid=0}^{\rm d}=\ket {\kd \xd}^{\rm d},
\eeq
at time $t$, where $X:=\int_0^t z(s)ds$ is the total accumulated noise. This phase noise is unknown to the experimenters, but it can be approximately corrected once we obtain its estimate, denoted by $c(Y)$. This is a function of all the results, obtained from measuring the SQ which for now we simply denote as $Y$. The correction is done via the control Hamiltonian, $\hat H_{\rm ctrl}^{\rm d} = - c(Y) \hat{\sigma}_z^{\rm d}/2$, which implements an instantaneous rotation around the $z$-axis by an angle $c(Y)$.

As described in Eq.~\eqref{eq:totalHamiltonian}, the SQ also senses the same noise as the DQ, but with a larger sensitivity $\ks$. Therefore we expect to be able to measure the SQ to obtain useful information about the noise $X$ and thus the DQ's phase error $\phi(X)$ as in Eq.~\eqref{eq:dataphase}. To simplify the process, we set the state of the SQ at the initial time, and immediately after each measurement, as the zero-phase state, $\ket{\phis=0}^{\rm s}$. Thus the integrated noise over time intervals between two measurements can be independently probed. From any time $t$ (immediately after the last measurement) to the time $t+\tau$ (immediately before the next measurement), the SQ's phase   becomes
\begin{equation}\label{eq:SQphase}
    \phis(\ddt) = K \!\! \int_{t}^{t+\tau} \!\! {\rm d}s ~ z(s) := K x.
\end{equation}
Here $x$ is the integrated noise during the duration $\ddt$, from $t$ to $t+\tau$. We assume the measurement on SQ can be done instantaneously with an observable of the form:
\begin{align}\label{eq:thetaop}
    \hat{\theta} := { \mathbb{1}}- \ket{\theta}^{\rm s}\bra{\theta}.
\end{align}
Here $\theta$ is a chosen measurement angle, 
\begin{align}
    \ket{\theta}^{\rm s} = (1/\sqrt{2}) \left(\ket{+1}_z^{\rm s} + e^{i \theta} \ket{-1}_z^{\rm s} \right),
\end{align}
and ${ \mathbb{1}}= \ket{+1}_z^{\rm s}\bra{+1} + \ket{-1}_z^{\rm s}\bra{-1}$ is the identity operator. The outcomes of the measurement, $y\in\{0,1\}$, are associated with projecting the SQ's state onto the two eigenstates, $\{\ket{\theta}^{\rm s},\ket{\theta+\pi}^{\rm s}\}$, respectively. 
Given the SQ's state, $\ket{\phis(\ddt)}^{\rm s} = \ket{K x}^{\rm s}$,  Born's rule leads to a probability of an outcome $y$ as 
\begin{equation} \label{eq:forwardP}
      \wp(y|\theta, x) := \big| \, ^{\rm s}\langle \theta+\pi \, y | \ks x \rangle^{\rm s}\, \big|^2 =  y + (-1)^{y}\cos^2\sq{\half (\theta - \ks x ) },
\end{equation}
which we called the likelihood function in \cite{song2023optimized}.

\subsection{Data qubit's coherence}\label{sec-dqcoh}

The natural quality measure we optimize is the averaged coherence of the data qubit with respect to the random noise and the phase correction. We denote the random variable, $\noise$, as a set of variables including both the noise and the phase correction. The definition of the coherence of the state by its phase is:
\begin{equation}\label{eq:coherence}
\coh:=|\langle e^{i\phi(\noise)}\rangle_\noise|=\left|\int\!\! {\rm d}\noise \,\, e^{i\phi(\noise)} \wp(\noise) \right|.
\end{equation}

The form of coherence analytically calculated in Refs.~\cite{song2023optimized,tonekaboni2023greedy} were for two extreme cases: the ``no-control" and ``with control" cases. For the former, the data qubit evolves stochastically under the noise. Its phase can be calculated as, $\phi(\noise)=\phi(X)=\kd\xd$, which is simply proportional to the total accumulated RTP. The coherence in \eqref{eq:coherence} thus can be written as:
\begin{equation}\label{eq:oldcoh_noctrl}
\cohnc:=|\langle e^{i\phi(X)}\rangle_{X}|=\left|\int e^{i\kd X} \wp(X ) {\rm d} X\right|.
\end{equation}
For the ``with control" case, we can use the measurement information $Y$ on the SQ to compute a phase correction, $-c(\info)$, and apply it to the data qubit, so that $\phi(\noise) = \phi(X,\info) = \kd X -c(\info)$. The coherence of the phase-corrected (``with control") data qubit is
\begin{align} 
    \coh_{c(Y)} := \Big|\big\langle e^{i [\kd X - c(\info)]}\big\rangle_{X,Y} \Big|=&\bigg| \sum_{Y} e^{-i c(Y)} \wp(Y)\int \!\! \dd X \,  e^{i \kd X}    \wp(X|Y)  \bigg|,\label{eq:gencoh_ctrl} \\
     \le &\, \sum_{Y}  \wp(Y)\left| \int \!\! \dd X \,  e^{i \kd X}   \wp(X|Y) \right|,\label{eq:oldcoh_ctrl}
\end{align}
In the last line, the inequality becomes an equality when 
\beq\label{eq:optctrl}
c(Y) =\arg  \int \!\! \dd X \,  e^{i \kd X}  \wp(X|Y) = \arg \big\langle e^{i \kd X} \big\rangle_{X|Y}
\eeq
which maximizes the phase-corrected coherence $\coh_{c(\bullet)}$ to a simpler formula: 
\begin{align} \label{eq:maxcoh}
    \cohc := \sum_\info  \wp(\info)\left| \,  \big\langle e^{i \kd X} \big\rangle_{X|\info}\right|.
\end{align}
Here, following Refs.~\cite{song2023optimized,tonekaboni2023greedy}, the c superscript stands for control (as opposed to nc above). 
However, this formula for the phase-corrected coherence can only be used when the optimal correction $c(Y)$ in  Eq.~\eqref{eq:optctrl} is applied to the data qubit. We will see later in Section~\ref{sec:sop} that when the correction is not optimal due to lack of knowledge, the coherence can only be computed from 
Eq.~\eqref{eq:gencoh_ctrl}.

\subsection{Generalized map-based formalism for RTP}\label{sec-genmap}

In this work, we consider more general scenarios with imperfections, where during the total time $T$, there are periods when the SQ can probe the noise (with the readout obtained and used in the phase error estimation) and periods when the SQ cannot probe the noise (\eg, during dead time of the SQ). Note that if the SQ can sense the noise but no measurement was made, it is equivalent to not probing. In this subsection, we extend the map-based formalism to be able to capture these scenarios. 

Let us now assume the classical noise $z(t)$ to be a RTP noise, which is a two-state fluctuator whose values can, without loss of generality, be $\pm 1$. In this case, the coherence Eq.~\eqref{eq:oldcoh_noctrl} and \eqref{eq:oldcoh_ctrl} can be calculated through two important mapping matrices $\mathbf{H}$ and $\mathbf{F}$ 
introduced in Refs.~\cite{tonekaboni2023greedy,song2023optimized}. These two maps can be seen as two extreme examples of the generalized case we consider here, which will be used 
in analyzing the SQ's measurement imperfection in Sec.~\ref{sec:sqreset}.

We use $\gu$ and $\gd$ to represent the rates of RTP jumping from $\rtp=-1$ to $\rtp = +1$ and from $\rtp=+1$ to $-1$, respectively. Given these, one can solve the master equation: 
\beq  \label{MESS}
\ul{\dot{P}}_t = 
\left( \begin{matrix} -\gd  & +\gu  \\ +\gd & -\gu \end{matrix} \right) \ul{P}_t,
\eeq
for probabilities of $\rtp$ at any given time, $\underline{P}_t = \left(\wp(\rtp_t = +1), \wp(\rtp_t=-1)\right)\!\tp$. This gives a steady state as $\ul{P}_{\rm ss} = \frac{1}{2 \mg}(\gu , \gd)\tp$, where $\mg = (\gu+\gd)/2$.
The fact that noise jumps between two values $z(t) \in \{ +1 , -1\}$ leads to the convenience in writing the coherence in terms of multiplication of $2\times 2$ matrices as in the following.

For a pedagogical purpose, let us consider an example with four time intervals marked by $t_0, t_1, t_2, t_3, t_4$, where the SQ only probes the noise during the intervals of duration $\tau_2 = t_2-t_1$ and $\tau_3 = t_3- t_2$, giving two readouts $y_2$ and $y_3$ at times $t_2$ and $t_3$, respectively. For the other two intervals, $\tau_1 = t_1-t_0$ and $\tau_4 = t_4 - t_3$, the SQ is either not probing the noise or no measurement on the SQ was made. Given this example, we first define the accumulated noises in each individual time interval as $x_1, x_2,..., x_4$, where
\beq
x_i = \int_{t_{i-1}}^{t_i} z(s) {\rm d}s,
\eeq
and compute the control coherence similar to Eq.~\eqref{eq:oldcoh_ctrl}. The total accumulated noise until the final time is a $1$-norm $X_4 \equiv ||\vec{x}_4||_1 = x_1 +x_2 + x_3 + x_4$ and $Y = \{ y_2, y_3\}$. The control coherence at time $t_4$ then becomes
\begin{align}\label{eq:excoh}
\coh_{c(Y)}:=& \bigg| \sum_{y_2, y_3} e^{-i c(Y)} \!\! \int \!\! \dd X_4 \, e^{i \kd X_4}   \wp(Y) \wp(X_4|Y)  \bigg|.
\end{align}
If the control is chosen to be the optimal one as in Eq.~\eqref{eq:optctrl}, then we have
\begin{align}\label{eq:excoh2}
\cohc =&  \sum_{y_2, y_3} \bigg|\int \!\! \dd X_4 \, e^{i \kd X_4}   \wp(Y) \wp(X_4|Y)  \bigg|= \sum_{y_2, y_3}\left| \sum_{z_0,..., z_4} \int \!\! {\rm d} x_1 \cdots {\rm d}x_4 \, e^{i \kappa ||\vec{x}_4||_1} \, \wp(\cdots)\right|,
\end{align}
where we have defined $z_n = z(t_n)$. In the second line, we have applied convolution (see Appendices in Ref.~\cite{tonekaboni2023greedy}) replacing $X_4$ with a summation $||\vec{x}_4||_1 = x_1 + x_2 + x_3 + x_4$ and replacing the integral measure $\int \!\! {\rm d}X_4$ with $\int \!\! {\rm d}x_1 \cdots {\rm d}x_4$, while the probability function becomes
\begin{align}\label{eq:exprob}
\wp(\cdots) \equiv \,\, &\wp(x_4,...,x_1, y_3, y_2, z_4,...,z_0) \nonumber\\
=\,\, &\wp(x_4, z_4 | z_3) 
\wp(y_3|x_3)\wp(x_3, z_3|z_2) \wp(y_2|x_2)\wp(x_2, z_2|z_1)\wp(x_1, z_1|z_0)\wp(z_0),
\end{align}
noting that $\wp(y_n|x_n)$ only appears for $n = 2$ and $3$. By substituting Eq.~\eqref{eq:exprob} back to Eq.~\eqref{eq:excoh2}, we can see that the integrals of $x_1$, ..., $x_4$ can be separated, but still with the correct ordering. Let us define the two integrals:
\begin{equation}\label{eq:H_element_def}
H_{z_{n-1}}^{z_n}(\tau_n,\kappa) := \!\!\int \!\! \dd x_n \, e^{i\kappa x_n}\, \wp(x_n,z_n|z_{n-1}) 
\end{equation}
for those $n$ with no measurement and
\begin{align}
\label{eq:F_elements_def}
    F_{z_{n-1}}^{z_n}\!\left(\mu_n, y_n\right) :=\!\!
    \int\!\! {\rm d}x_n \, e^{i \kd x_n}\,\wp_{\meas_n}(y_n|x_n)\,\wp(x_n,z_n|z_{n-1}),
\end{align} 
for those $n$ with measurement results $y_n$. The new variable $\mu_n$ is a measurement strategy, $\mu_n=\{\theta_n,\ddt_n\}$, which includes the measurement angle $\theta_n$ and the waiting time $\ddt_n$ for the $n$-th measurement. Using Eq.~\eqref{eq:H_element_def} and \eqref{eq:F_elements_def}, we can simplify the integrand of the coherence in Eq.~\eqref{eq:excoh2} to
\begin{align}
\int \!  {\rm d}x_4  \cdots {\rm d}x_1 \, e^{i \kappa ||\vec{x}_4||_1} \, \wp(\cdots)=  H^{z_4}_{z_3}(\tau_4,\kappa) F^{z_3}_{z_2}(\mu_3,y_3) F^{z_2}_{z_1}(\mu_2,y_2) H^{z_1}_{z_0}(\tau_1,\kappa) \wp(z_0). \nonumber
\end{align}
Then, with the summation of $z_0,...,z_4$ in Eq.~\eqref{eq:excoh2}, the expected coherence can be formulated into a matrix-multiplication as
\begin{align}
\cohc = \!\! \sum_{y_2, y_3} \left| \ul{I}^\top {\bf H}(\tau_4,\kappa) {\bf F}(\mu_3,y_3) {\bf F}(\mu_2,y_2) {\bf H}(\tau_1,\kappa) \ul{P}_0 \right|,
\end{align}
where $\ul{P}_0 = (\wp(z_0 = +1), \wp(z_0 = -1))\tp$ and $\ul{I}^\top = (1,1)$. One might notice that there are four matrices, where ${\bf H}$ is used for the intervals without measurement and ${\bf F}$ is used for those with measurement information. These matrices can be computed explicitly~\cite{tonekaboni2023greedy}. For ${\bf H}$,
\begin{equation}\label{eq:Hmatrix}
{\bf H}(\tau,\kappa):= \left( \begin{matrix}H^{+1}_{+1}(\tau,\kappa) &  H^{+1}_{-1}(\tau,\kappa) \\[5pt] H^{-1}_{+1}(\tau,\kappa) &  H^{-1}_{-1}(\tau,\kappa) \end{matrix} \right),
\end{equation}
we have
\begin{equation}
H_{z_{n-1}}^{z_n} = {\rm exp}(-\bar{\gamma} \ddt)\begin{cases}
\cosh\left(\dfrac{\lambda}{2}  \ddt \right)- s \dfrac{\eta}{\lambda}\sinh\left(\dfrac{\lambda }{2} \ddt\right),& \text{for } z_{n}=z_{n-1} = s
\\ 
\dfrac{2 \gu}{\lambda} \sinh\left(\dfrac{\lambda}{2} \ddt\right),& \text{for } z_{n}=-z_{n-1} = +1
\\
\dfrac{2 \gd}{\lambda} \sinh\left(\dfrac{\lambda}{2} \ddt\right),& \text{for } z_{n}=-z_{n-1} = -1
\end{cases},
\end{equation}
using $s \in \{ -1, +1\}$ and
\begin{equation}
    \begin{aligned}
      \lambda(k)&=\sqrt{(\gd+\gu)^2-4i k (\gd-\gu) -4 k^2},\\
\eta(k)&=(\gd-\gu) -2i k.  
    \end{aligned}
\end{equation}
For the $\mathbf{F}$ matrix, 
\begin{equation}\label{eq:Fmatrix}
{\bf F}\left(\mu_n, y_n\right):= 
\left( 
\begin{matrix}
F^{+1}_{+1}\left(\mu_n, y_n\right) &  F^{+1}_{-1}\left(\mu_n, y_n\right) \\[5pt] 
F^{-1}_{+1}\left(\mu_n, y_n\right) & F^{-1}_{-1}\left(\mu_n, y_n\right) 
\end{matrix} 
\right),
\end{equation}
its elements can be computed explicitly from ${\bf H}$ as:
\begin{align}
\label{eq:FH}
\ffun\left(\meas{=}\{\theta,\ddt\},y\right)  = & \frac{1}{4}\Big[2 \, \hfunc(\ddt,\kd) +(-1)^{y}e^{-i\theta} \hfunc(\ddt,\kd+\ks) +(-1)^{y}e^{+i\theta}\hfunc(\ddt,\kd-\ks)\Big].
\end{align}

In summary, the two mapping matrices ${\bf H}$ and ${\bf F}$, can be combined in specific ways to calculate the DQ's coherence for situations where there are time periods when
the SQ can probe and cannot probe the noise. We can define a coherence vector at time $t_n$ as
\begin{align}
\ul{A}_n = {\bf M}_n {\bf M}_{n-1} \cdots {\bf M}_2\, {\bf M}_1 \ul{A}_0
\end{align}
where ${\bf M}_n$ can be ${\bf H}(\tau_n, \kappa)$ or ${\bf F}(\mu_n,y_n)$ and we set $\ul{A}_0 = \ul{P}_0$. The coherence is then reduced to
\begin{align}
\coh_{c(Y)} = \left| \sum_{Y} e^{-i c(Y)} \ul{I}^\top \ul{A}_n \right| \le \sum_Y \left| \ul{I}\tp \ul{A}_n \right| = \cohc,
\end{align}
\ie, it has a maximum at $\cohc$ when the control is optimal as $c(Y) = \arg \ul{I}\tp \ul{A}_n$.

Using the above formulation, the two extreme cases considered in the previous work \cite{tonekaboni2023greedy,song2023optimized} can be re-derived. For the no control case with no information from the SQ, we can use the total time $T$ as one interval with the total accumulated noise $X$. The dynamics is described by one single $\bf H$ map, which gives the coherence Eq.~\eqref{eq:oldcoh_noctrl} as
\begin{align}\label{eq:Coh_nc_A}
    \cohnc &:= \left| \left\langle e^{i \phi(\xd)} \right\rangle_\xd \right|= \left|\underline{I}^\top 
    {\bf H}(T,\kappa) \underline{A}_0 \right|.
\end{align}
In an asymptotic regime where $\kappa\ll\mg\ll K$, it was shown in \cite{tonekaboni2023greedy} that the coherence under no control shows exponential decay $
\cohnc(T)  \sim \exp(- \Gamma^{\rm nc} T)$, where 
\begin{align}\label{eq:gammanc}
    \Gamma^{\rm nc} = \kd^2 \bg/2 \mg^2
\end{align}
is the decoherence rate and $\bg \equiv 2\gu\gd/(\gu+\gd)$. 

For the case with (continual) measurement, the SQ is measured at times $t_1, t_2,...,t_{N} = T$ to probe every individual part of the accumulated noises $x_1, x_2,..., x_N$, giving a string of readouts $Y_N = \{ y_1, y_2,..., y_N \}$. The coherence is improved through a series of mapping matrices ${\bf F}$ with the measurement information $Y_N$ as:
\begin{align}\label{eq-fullmeascoh}
    \coh_{c(Y_N)} 
    &= \bigg| \sum_{\info_N} e^{-ic(\info_N)} \ul{I}\tp \ul{A}_N \bigg| \le \sum_{\info_N}\big |  \ul{I}^\top\,  \ul{A}_N \big|,
\end{align}
where $\ul{A}_N = {\bf F}(\meas_N, y_N)\,\cdots  {\bf F}(\meas_2,y_2) {\bf F}(\meas_1,y_1) \underline{A}_0$. As before, the coherence is maximized when the final phase correction is chosen to be optimal $c(\info_N)={\rm arg}\, \ul{I}\tp \ul{A}_N$.

\subsection{Adaptive measurement and control strategy} 
\label{sec:prederiveH}

It has been shown in previous work \cite{tonekaboni2023greedy,song2023optimized} that, in the asymptotic regime, where $\kappa\ll \gu,\gd\ll \ks$, and in the long-time limit, the data qubit decoherence rate can be improved relative to the no-control coherence in Eq.~\eqref{eq:Coh_nc_A} by a multiplicative factor of $1.254(\mg/\ks)^2$, using an adaptive algorithm called MOAAAR, which stands for Map-based Optimized Adaptive Algorithm for Asymptotic Regime. 
This MOAAAR algorithm sets the SQ's measurement angles and the waiting times according to the following adaptive relation:
\begin{subequations}\label{eq:adtmo}
\begin{equation}\label{eq:adtmoa}
\theta_{n+1}= s_n \Theta,
\end{equation}
\begin{equation}\label{eq:adtmob}
\ddt_{n+1}=\Theta/\ks,  
\end{equation}
\end{subequations}
where $s_n \in \{ +1, -1\}$ is adaptively chosen depending on the sign of the quantity:
\begin{equation} \label{defzeta}
    \zeta_n :=\frac{|A_n^{\rtp=+1}|-|A_n^{\rtp=-1}|}
    {|A_n^{\rtp=+1}|+|A_n^{\rtp=-1}|},
\end{equation}
which is the approximated mean of $z_n$ conditioned on $Y_n$, the information available up to and including the time $t_n$. Here, $A_n^{\rtp=+1}$ and $A_n^{\rtp=-1}$ are the first and the second element in the vector $\ul{A}_n$, respectively. The variable $s_n$ can also be interpreted as the most likely value of $z_n$, conditioned on $Y_n$. There is another statistically important quantity that will also be used later:
\begin{equation}\label{defalpha}
    \alpha_n:=\frac{K}{\kappa}\arg\frac{A_n^{z=+1}}{A_n^{z=-1}}.    
\end{equation}
This can be interpreted as a variable of order unity which is proportional to the maximum difference of an optimal DQ's control phase from hypothetically finding out if the RTP state was $z_n =+1$ or $z_n=-1$. The variables $\alpha_n$ and $\zeta_n$ are the two sufficient statistics that can be computed from the coherence vector $\ul{A}_{n}$ given the string of measurement records $Y_n$ in the past. 

\begin{figure}[t]
    \centering    \includegraphics[width=0.7\textwidth]{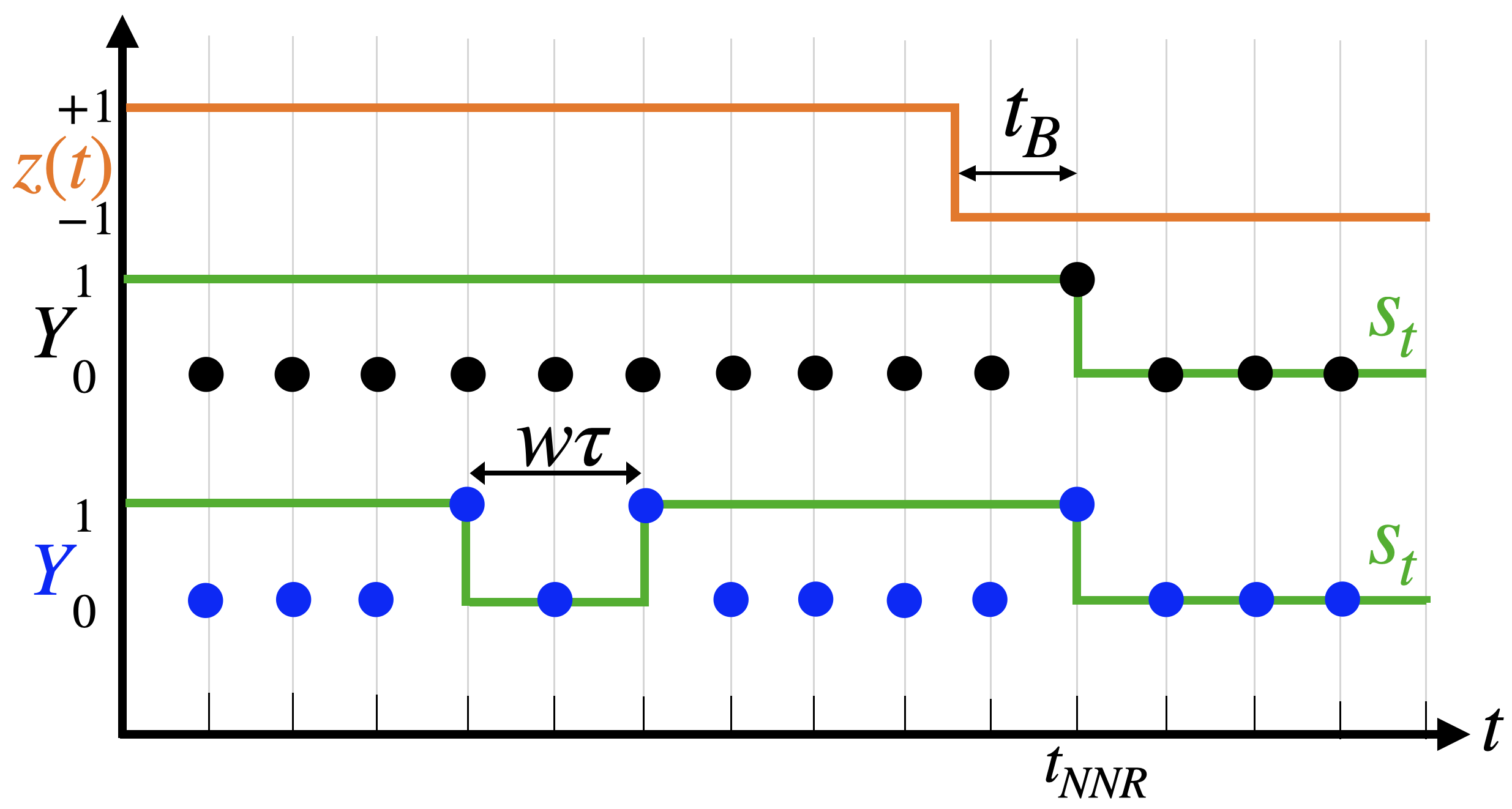}
    \caption{Schematic pictures to illustrate a Before-time ($t_{\rm B}$) and a time between a pair of NNRs ($w \tau$), which will be used in analytical calculations of coherence in Section~\ref{sec:rederiveH} and Section~\ref{sec:delthe}. In the first row, we show a realization of the RTP, $z(t)$, with one flip from $+1$ to $-1$ (orange line). The second row shows possible SQ's measurement results for a scheme with no imperfection (black dots), where one NNR is observed at $t_{\rm NNR}$. This observation indicates a detection of the jump in $z(t)$, which leads to the flip of $s(t)$ (green line), \ie, the observer's guess of $z(t)$. Here $s(t)$ is related to $s_n$, the sign of $\zeta_n$ in Eq.~\eqref{defzeta}, by $s(t)=s_n$ for $t\in(t_n,t_{n+1}]$. Thus, the Before-time, $t_{\rm B}$, is the time difference between the real flip time of $z(t)$ and the detected flip of $s(t)$. The last row shows possible measurement results when there are imperfections $\delthe \ne 0$ or $\measerr \ne 0$, where we have denoted $w\ddt$ to represent the time between a pair of false NNRs, wrongly indicating that $z(t)$ had jumped twice, as seen by the two consecutive wrong jumps in $s_n$.}
    \label{fig:tBandw}
\end{figure}

To understand this adaptive strategy in simple terms, it is an algorithm that will give null measurement results $y_n = 0$ whenever the value of $s_n$ is a correct estimate of $z_n$ and $z_n$ does not flip, and will give a non-null result (NNR) whenever $z_n$ flips, thereby detecting the RTP flip effectively. To see this, let us assume that, for any $n$, we have guessed $s_n = z_n$ correctly. Then, if $z_n$ does not change during the SQ probing time, the accumulated noise for the next measurement is exactly $x_{n+1} = z_n \tau_{n+1}$. Then, with the MOAAAR's measurement angle and time as in Eq.~\eqref{eq:adtmo}, one can show that the probability of getting a NNR,
\begin{align}\label{eq-probone}
    \wp(y_{n+1} = 1|\theta_{n+1}, x_{n+1}) =  1- \cos^2(0) = 0,
\end{align}
is exactly zero, by substituting $\theta_{n+1} = s_n \Theta$ and $x_{n+1} = z_n \Theta/K$ in Eq.~\eqref{eq:forwardP}. Therefore, as long as the RTP stays unchanged, the SQ's measurement results will remain as null results to give a sequence of records $Y_n = \{ 0,0,...,0\}$. When a flip occurs, $z_{n+1} \ne z_n$, then $\wp(y_{n+1}=0| \theta_{n+1},x_{n+1})$ becomes smaller and getting the NNR $y_{n+1} = 1$ is now possib: the exact probabilities depend on when during $[t_n,t_{n+1}]$ the RTP jumps. This means that, for MOAAAR, getting a NNR indicates that the RTP has flipped with a high probability. After that, if $z_n$ stays unchanged, the null result will confirm that. With the relation in \eqref{eq:adtmo}, the MOAAAR algorithm means that the matrix $\bf{F}$ can be treated as a single-parameter map, $\ffun_{s}^{y}(\Theta):=\ffun(s \Theta, \Theta/K, y)$. Moreover, it was shown semi-analytically, with the aid of \emph{Mathematica} that $\Theta = \Theta^* \approx 1.50055$ gives the lowest decoherence rate.

\subsection{``Before-time" analytical calculation}\label{sec:rederiveH}

In this subsection, we will re-derive the DQ's coherence as a result of the MOAAAR strategy obtained in the previous work, using a much more elementary and intuitive analysis. It is based on the ``Before-time", $t_{\rm B}$, which is defined as the delay between the real RTP flip and the detected flip by $s_n$, associated with the subsequent NNR $y(t_{\rm NNR}) = 1$. The Before-time $t_{\rm B}$ is illustrated in Fig.~\ref{fig:tBandw}, where the orange line (top example) is the real RTP and the green line (middle example) is the estimated RTP denoted by $s_t$. Then, $t_{\rm B}$ is the time between the real flip and the detected flip. This method of calculation using the Before-time will be fruitful when we analyze the imperfection from the measurement angle, $\delthe$, in later sections. In order to compute the coherence at some time $T_1$, in the asymptotic limit, we can consider that the RTP sporadically flips, once in a while, and the SQ will detect each flip of the RTP by obtaining a NNR at some Before-time later.

Consider a single NNR, as in Fig.~\ref{fig:tBandw}, occurring at $T_1 = t_{\rm NNR} = N_1 \tau$. We calculate the coherence as in Eq.~\eqref{eq:oldcoh_ctrl} which is simply an absolute value of the averaged phase difference between the real DQ's phase, $\phi = \kappa X$ and the phase estimation $\varphi = c(Y)$. When the decoherence is small, it equals half the mean square error (MSE) in the final phase estimate \cite{tonekaboni2023greedy}, \ie,
\begin{align}\label{eq-beftimecoh}
\cohc\approx 1 - \frac{1}{2} \left\langle [\phi(T_1) - \varphi_{T_1}]^2 \right\rangle ,
\end{align}
where the expectation average is over all possible random variables. Let us first assume $z(t=0) = 1$ as in the orange curve of Fig.~\ref{fig:tBandw}. The real phase of the DQ can be calculated directly from $t_{\rm B}$, which gives
\begin{align}\label{eq:reDQphase}
\phid(T_1)=\kappa \, (T_1-t_{\rm B})(+1)+\kappa \, t_{\rm B}(-1) =\kappa \, T_1-2\kappa \, t_{\rm B},
\end{align}
where the first and second terms are the real-phase contributions from the time duration when $z(t) = +1$ and $z(t) = -1$, respectively. The estimated phase, $\varphi$ can also be obtained,
\beq\label{eq:esDQphase}
\varphi_{T_1}=\kappa T_1-2\kappa\langle t_{\rm B}\rangle,
\eeq
which is simply the unbiased estimator of $\phid(T_1)$. 
We also show in \ref{sec:appendix1} that the phase estimator $\varphi_{T_1}$ can be obtained explicitly, confirming that the result above is correct up to $\mathcal{O}(\mg^2/\ks^2)$, which can be ignored in the asymptotic regime. 
The MSE in Eq.~\eqref{eq-beftimecoh} thus becomes
\beq\label{eq:decoMSE}
\begin{aligned}
\langle[ \varphi_{T_1}-\phid(T_1)]^2 \rangle_{t_{\rm B}}=4\kd^2\left(\langle t_{\rm B}^2\rangle-\langle t_{\rm B} \rangle^2\right),
\end{aligned}
\eeq
where the expectation average has changed to over the random variable $t_{\rm B}$. 
We then have to compute the probability distribution of $t_{\rm B}$ and its moments.

The probability distribution of $t_{\rm B}$ can be obtained from the probability of $Y_n$ given $t_{\rm B}$ using Bayes' theorem. Within the time duration $T_1 = N_1 \ddt$, the string of measurement results is $Y_{N_1} = \{ 0,0,\cdots,0,1\}$, where the NNR, $y_n = 1$ occurs at $t_{\rm NNR}$. For convenience, we 
divide the total interval into the following sub-intervals:
\beq
[0,T_1) = \bigcup_{m=0}^{N_1 -1} [m \tau, m\tau + \tau),
\eeq
and define 
a Boolean variable 
${\cal T}_m$ that is true if and only if $t_{\rm B}$ falls within the $m$th interval, \ie,
\beq 
{\cal T}_m \equiv\left\{
\begin{aligned}
1,& \text{ if } t_B\in[m\tau, m\tau+\tau)\\
0,& \text{ otherwise. } 
\end{aligned}
\right. 
\eeq 
This Boolean variable works as a flag, i.e., when it is used together with $t_B$, it implies that $t_B$ does fall within the $m$th interval. 


We then calculate the probability of $t_{\rm B}$ by using Bayes' theorem, starting with the probability that is in each subinterval,
\begin{align}\label{eq-probtbfull}
    \wp(Y_{N_1},t_{\rm B}|{\cal T}_m) = &\, \wp(Y_{N_1}|t_{\rm B},{\cal T}_m ) \, \wp(t_{\rm B}|{\cal T}_m )=\, \wp(Y_{N_1} |t_{\rm B},{\cal T}_m)(1/\tau),
\end{align}
where we have used $\wp(t_{\rm B}|{\cal T}_m) = 1/\tau $, a uniform distribution of $t_{\rm B}$ in the $m$th $\tau$-duration  interval, because of the constant transition rate.

Therefore, we need the probability of the NNR measurement result given $t_{\rm B}$ in each interval, $\wp(Y_{N_1} | t_{\rm B},{\cal T}_m)$. A special case is $m = 0$, when $t_{\rm B}$ is in the subinterval right before the NNR. In this case, the real RTP stays $+1$ for all the subintervals associated with all null results. That means, $s_n = z_n$ and the unit probability Eq.~\eqref{eq-probone} is applied for all measurement results, except the last $y_{N_1} = 1$. This gives
\beq\label{eqn:probtB0}
\begin{aligned}
\wp(Y_{N_1}|t_{\rm B},{\cal T}_0)
= \,\, \wp(y_{N_1} = 1|x=\ddt-2t_{\rm B})= \,\, 1-\cos^2\left(\ks t_{\rm B}\right),
\end{aligned}
\eeq 
which is simply the probability of NNR $y_{N_1}=1$ conditioned on the accumulated noise $x = t_{\rm B} (-1) + (\tau - t_{\rm B})(+1)$ during that sub-interval.

\begin{figure}
    \centering
    \includegraphics[width=0.7\textwidth]{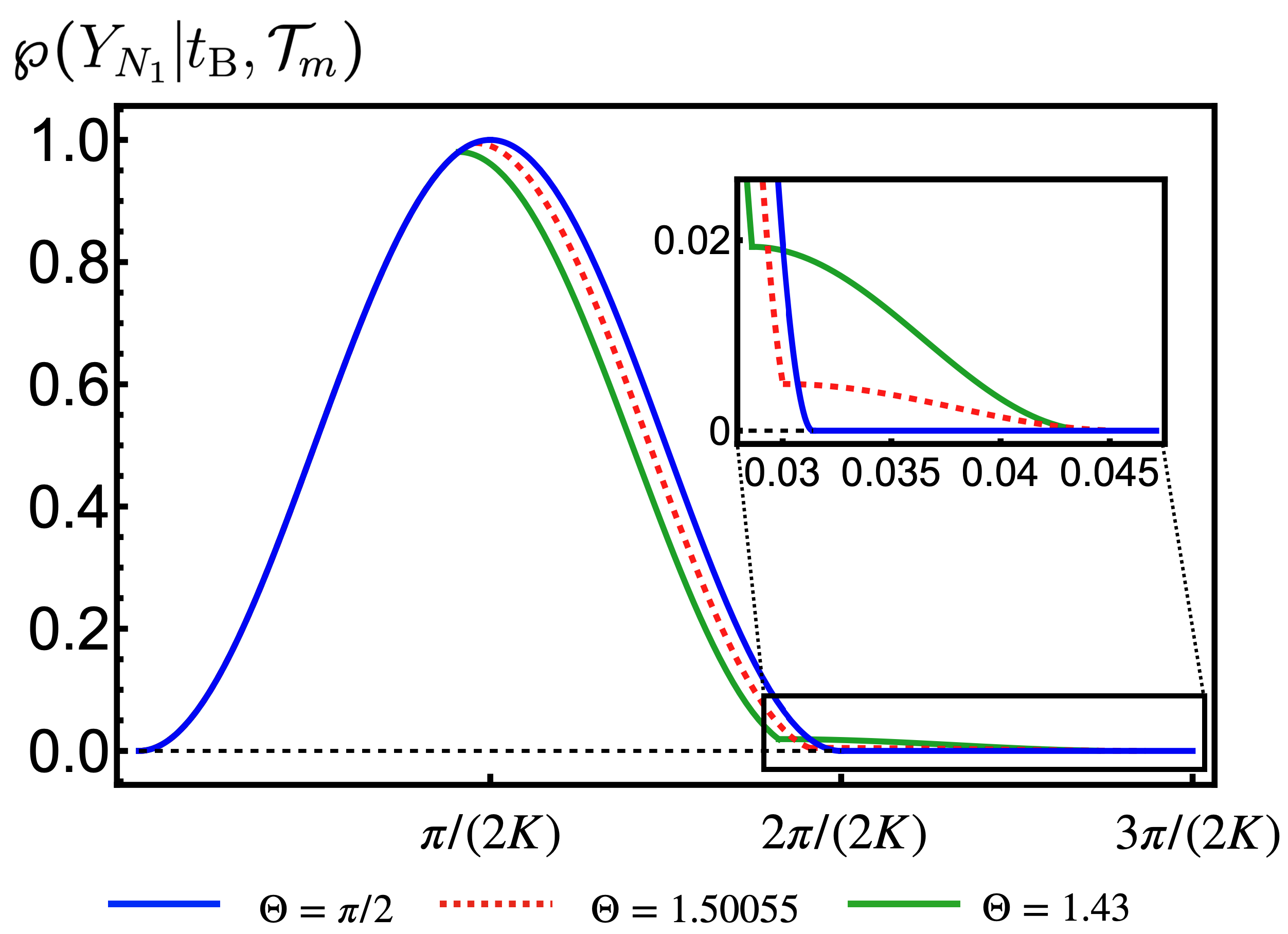}
    \caption{Unnormalized probability functions, $\wp(Y_{N_1} |t_{\rm B},{\cal T}_m)$ for $m=0,1,2$ in Eq.~\eqref{eq-normedprobtb}, are plotted as a function of $t_{\rm B}$, where $t_{\rm B}$ (Before-time) is the time between the actual flip and the detected flip of the RTP. Here $\tau$, the measurement waiting time, is different for the three different curves, according to $\tau = \Theta/K$. The ticks of the horizontal axis are chosen to be multiples of $\pi/(2K)$, which is the value of the optimal measurement waiting time $\tau$ for $\Theta = \pi/2$. The three values of $\Theta$ chosen here are $\Theta = \pi/2 \approx 1.57$ (blue), $\Theta = \Theta\opt \approx 1.50$ (dashed red), and $\Theta = 1.43$ (green), all with $\ks=100$. The main plots show that the highest probability for $t_B$ occurs at $t_{\rm B} = \Theta/K$, while the probabilities are very small when $t_{\rm B} > 2\Theta/K$. The zoomed-in plots (inset) for the latter region show that, even though the probability is exactly zero for $\Theta=\pi/2$, for other values of $\Theta$, the probabilities in that range are finite, with kinks at $t_{\rm B} = 2 \Theta/K$. The black dashed lines mark where the ordinate is zero.}
    \label{fig:disoftBforthree}
\end{figure}

For other $m \ne 0$, there are non-unity contributions to $\wp(Y_{N_1}|t_{\rm B},{\cal T}_m)$ which are from: (a) a null result at the end of the subinterval $[m\tau,m\tau+\tau)$ where the RTP flips, (b) a series of $m-1$ null results after (a) from that $z_n \ne s_n$ (the RTP flip, but the measurement result is still null), and (c) a NNR when $z_n \ne s_n$. This gives
\begin{align}\label{eqn:probtBm}
\wp(Y_{N_1}|t_{\rm B}, {\cal T}_m) = &\,\, \wp(y_{(a)}=0|x_{(a)}=2m\ddt + \ddt -2t_{\rm B}) \times \left[\wp(y_{(b)}=0|x_{(b)}=-\ddt)\right]^{m-1}\nonumber \\
& \times \wp(y_{(c)}=1|x_{(c)}=-\ddt)\nonumber \\
&=\cos^2(-m\Theta+\ks t_{\rm B})\left(\cos\Theta\right)^{2(m-1)}(1-\cos^2\Theta).
\end{align}
We substitute this result back in Eq.~\eqref{eq-probtbfull} and then sum over all interval $[m\tau,m\tau+\tau)$ from $m = 0$ to $N_1 -1$ to obtain the probability of $t_{\rm B}$,

\begin{align}\label{eq-normedprobtb}
    \wp(Y_{N_1},t_{\rm B}) = &\, \sum_{m=0}^{N_1-1} \wp(Y_{N_1},t_{\rm B}|{\cal T}_m) \wp({\cal T}_m) = \,\frac{1}{\tau}\sum_{m=0}^{N_1-1}  \wp(Y_{N_1} |t_{\rm B},{\cal T}_m).
\end{align}
Here, $\wp({\cal T}_m)=1$ when $t_B\in[m\tau, m\tau+\tau)$ and $\wp({\cal T}_m)=0$, otherwise. We plot $\wp(Y_{N_1} |t_{\rm B},{\cal T}_m)$ for $m=0,1,2$ in Fig.~\ref{fig:disoftBforthree}. We only plot three intervals for different $\Theta$, since the probabilities of $t_{\rm B}$ in the subsequent intervals are too small to be visible.

We can now use Eq.~\eqref{eq-normedprobtb} to get the probability density $\wp(t_{\rm B})$ of $t_{\rm B}$ by taking the limit $N_1 \rightarrow \infty$ in $\wp(Y_{N_1},t_{\rm B})$, since $t_B$ is only defined when we have $Y_{N_1}$ that means when we have one NNR. Note that this limit is mathematical rather than physical --- in reality there is not an infinite time $T_1$ before an isolated NNR in which the jump could have taken place because there would have been another NNR at some earlier time. However, as shown in Fig.~\ref{fig:disoftBforthree}, when $\Theta \approx \pi/2$, the jump will almost certainly have occurred in the intervals labelled $m=0$, $m=1$, or $m=2$. Thus taking $N_1=3$ would give results almost identical to $N_1\to\infty$, but it is only in the latter limit that the $\wp(t_{\rm B})$ as defined above integrates exactly to unity. We verify this with the help of \emph{Mathematica}, and then calculate the mean value, 
\begin{equation}\label{eq:meanoftB}
    \langle t_{\rm B}\rangle=\frac{1} {2\ks} \left(\Theta+\cot \Theta+\Theta\csc^2 \Theta \right),
\end{equation}
and the variance,
\begin{align}\label{eq:varoftB}
\langle t_{\rm B}^2\rangle-\langle t_{\rm B}\rangle^2 = \,\, \frac{1}{4 \ks^2}\Big[3\Theta^2 \csc^4 \Theta+ \tfrac{1}{3}\Theta^2  - 1 - [2\Theta (\Theta - \cot\Theta) +1] {\rm csc}^2\Theta\Big].    
\end{align}
These moments are used to compute the mean square error in Eq.~\eqref{eq:decoMSE} and thus the coherence as in Eq.~\eqref{eq-beftimecoh}. However, we remind the reader that so far we only considered the case of the RTP starting from $z(t=0) = +1$ then jumping to $-1$, but not the opposite. 
Therefore, for an asymptotically long time $T_1$, if we assume independent identically occurrence of NNRs, then we multiply the reduction of coherence per single NNR: $1-\cohc_{c(Y)} = 2 \kd^2 \left( \langle t_{\rm B}^2 \rangle - \langle t_{\rm B} \rangle^2\right)$ with $\gd T_1$ or $\gu T_1$, \ie, the average numbers of NNRs occurred during time $T_1$ given $z(t=0) = +1$ or $-1$, respectively. Since we consider the asymptotic regime, we can use $\wp(z_t=+1)=\gu/(\gu+\gd)$ and $\wp(z_t=-1)=\gd/(\gu+\gd)$. That is, the average decoherence rate, defined as the coherence reduction divided by the total time $T_1$, $\bar{\Gamma}:=(1-\cohc_{c(Y)})/(\cohc_{c(Y)}T_1)$, in the asymtotic limit is given by 
\beq\label{eq:decoresult}
\begin{aligned}
\bar{\Gamma}&\approx \frac{1}{T_1} 2\kappa^2(\langle t_{\rm B}^2\rangle-\langle t_{\rm B}\rangle^2)\left(\gu T_1\frac{\gd}{\gu+\gd}+\gd T_1\frac{\gu}{\gu+\gd}\right)=\frac{\kappa^2\bg}{2\ks^2}H_\Theta,
\end{aligned}
\eeq
where $\bg \equiv 2\gu\gd/(\gu+\gd)$ and
\beq\label{eq:Hfunction}
H_{\Theta} := 
 3\Theta^2 \csc^4 \Theta - [2\Theta (\Theta - \cot\Theta) +1] {\rm csc}^2\Theta + \tfrac{1}{3}\Theta^2  - 1.
\eeq
This is exactly the same result found in ~\cite{tonekaboni2023greedy} by studying the eigenvalues of the ${\bf F}$ maps in the limit $\kappa \ll \ks$.  Through finding the minimum point of $H_\Theta$, we get a plausibly optimal MOAAAR algorithm with $\meas=\{s\Theta\opt, \Theta\opt\}$, which gives the decoherence rate $H\opt = H_{\Theta\opt}=1.254$ for $\Theta\opt\approx 1.50055$. 

As seen above, the concept of Before-time, $t_{\rm B}$, has offered a more elementary approach to re-derive $H_\Theta$ in Eq.~\eqref{eq:Hfunction} than that was done in the previous work~\cite{tonekaboni2023greedy,song2023optimized}. 
With the current approach, we also obtained the probability of $t_{\rm B}$ in Eq.~\eqref{eq-normedprobtb}, which can give an insight into why the optimal measurement angle, $\Theta\opt \approx 1.50055$ is slightly less than $\Theta = \pi/2$. Recall that the decoherence rate is directly proportional to the variance of $t_{\rm B}$. We can understand how the variance changes with $\Theta$ from Fig.~\ref{fig:disoftBforthree}, plotted for three values of measurement angles:  $\Theta = 1.43 < \Theta\opt$ (blue), $\Theta = \Theta\opt \approx 1.50055$ (dashed red), and $\Theta = \pi/2 > \Theta\opt$ (green). 

Considering the plots of probability over the range of $t_{\rm B} \in [0, \pi/K]$, one would guess that the variance of $t_{\rm B}$ should get smaller as $\Theta$ decreases. That is, the distribution gets narrower because one measures more often. However, if one considers the contribution from small probability in the range $t_{\rm B} >2\Theta/\ks $ (the inset of Fig.~\ref{fig:disoftBforthree}), then one sees that this causes the variance to get larger as $\Theta = K\tau$ decreases below  $\pi/2$. This is because, for $\tau < \pi/(2K)$, at the time of measurement, the state arising from there having been no RTP jump since the last detected one is not orthogonal to the state arising from there having been one RTP jump in the interval prior to the current one. This means that the NNR will not occur with probability one in the latter instance. Therefore, one jump of RTP cannot be perfectly detected in two intervals (two measurements), and the third interval is needed when $\Theta$ is close to $\pi/2$ (the inset of Fig.~\ref{fig:disoftBforthree}). Thus, the smaller waiting time (smaller $\Theta$) implies  the jump will most likely be detected earlier, but there is a chance it will take longer because a third interval is required (or, with very low probability, still further intervals).
Therefore, these are two competing contributions to the variance. The first decreases linearly with $\pi/2 - \Theta$ when this is small, while the latter increases quadratically in this. This explains why there is an optimum value, $\Theta\opt$, of $\Theta$, slightly less than $\pi/2$, where the variance of $t_{\rm B}$ (and hence the DQ decoherence rate) is the smallest.


Furthermore, we notice another interesting finding: the mean of $t_{\rm B}$ is linearly related to the slope of the estimated phase in the single measurement step just before we obtain $y_{n+1}=1$. Here the slope, $\lambda_{s_{n+1}}^{y_{n+1}}$ is defined as an always-positive number by the equation  $\lambda_{s_{n+1}}^{y_{n+1}}s_{n+1}\kd \tau=\varphi(t_{n+1}) - \varphi(t_{n})$. If we are in the asymptotic regime and we always obtain the null results, we can expect the slope in one measurement step of the estimated phase, $\lambda_{s_{n+1}}^{y_{n+1}}$, be $1$. Once we obtain a NNR, the slope is expected to be a constant that is less than but close to $1$. The detailed relation between $\langle t_{\rm B}\rangle$ and the slope can be found in \ref{sec:appendix1}. The concept of $t_{\rm B}$ will also be used later in Section~\ref{sec:delthe}, where we analyze the effect of the measurement angle uncertainty to the decoherence rate.


\section{Imperfection analysis}\label{sec:imperfections}
We will consider three categories of imperfections in this paper, labeled by `A', `B', and `C' in Figure \ref{fig1}. The first label `A' represents imperfections that arise in the measurement process of the SQ, including the measurement angle uncertainty $\delthe$, the SQ's readout and reset time $\tsr$, the detector dead time $\deadt$, and the measurement readout error $\measerr$. The label `B' denotes the uncertainty of the DQ's sensitivity to the RTP. Lastly, the label `C' indicates the additional dephasing of the SQ, which might be a result of other types of noise besides the considered RTP. We explain all these imperfections in detail as follows:

\begin{figure}[t]
    \centering
    \includegraphics[width=0.7\textwidth]{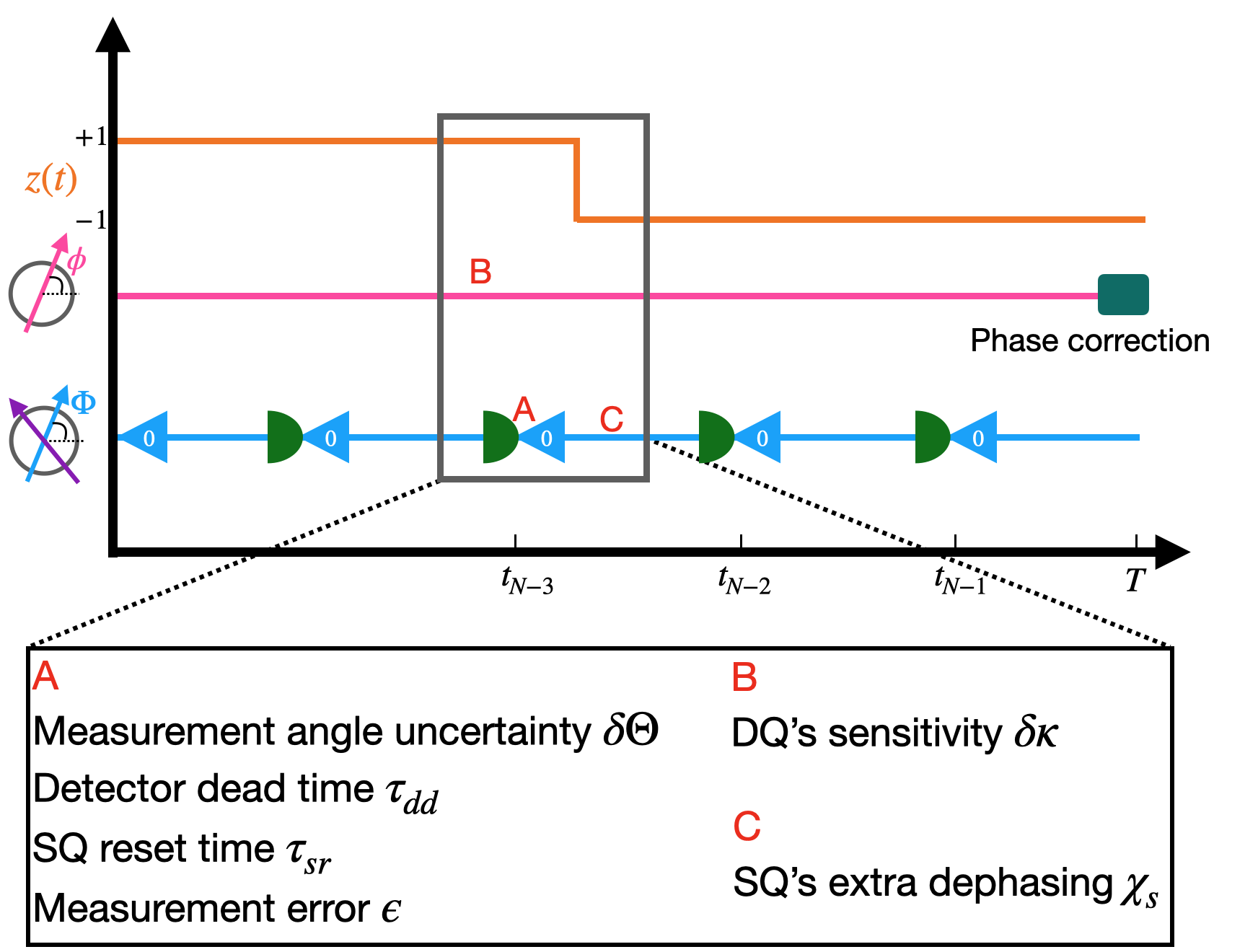}
    \caption{Schematic diagrams showing the SQ's measurement and decoherence mitigation with different types of imperfections we consider in this work. The imperfections we consider in this paper are labelled in the dark gray box with text explanations below.}
    \label{fig1}
\end{figure}

\textit{The uncertainty in measurement angles, $\delthe$} -- In a real experiment, even though a measurement angle is theoretically set at $\Theta$, there can be an unknown uncertainty resulting in an actual angle being shifted to $\Theta + \delthe$, which affects statistics of the SQ's measurement results.

\textit{The SQ's readout and reset time, $\tsr$} -- For each measurement, the SQ should be read out and then be reset to its zero-phase state. These processes do take time. During this readout and reset time, the SQ will not be able to sense the RTP, but the noise does still affect the DQ causing additional decoherence. 

\textit{The detector dead time, $\deadt$} -- This happens when any detection process of the SQ is not functioning, causing a dead time where the SQ cannot be measured. We will see that if the dead time is longer than the optimal measurement time, $\deadt \ge \tau$, then it is better to wait and measure the SQ at a suboptimal time instead of measuring the SQ immediately at $\deadt$ when the detector is ready.

\textit{The measurement error and additional dephasing} -- These two imperfections affect the coherence of the DQ in a similar way. The measurement error, $\measerr$, is the rate at which the SQ's measurement randomly gives wrong results, while the extra dephasing rate $\sqdepha$ is caused by other noise sources and environment on the SQ.

\textit{The uncertainty in the DQ's noise sensitivity, $\delkd$} -- Similarly to the uncertainty in the measurement angle, there can be uncertainty in the sensitivity, $\kd$, of the DQ. This imperfection affects the DQ's decoherence directly, but does not change the statistics of the SQ's measurement results.

\subsection{Coherence calculation with imperfection}\label{sec:sop}

In Section~\ref{sec-genmap}, we have emphasized that the coherence, \eg, in Eq.~\eqref{eq-fullmeascoh}, can be maximized only when the final phase correction is chosen to be optimal $c(\info_N)={\rm arg}\, \ul{I}\tp \ul{A}_N$. However, one has to realize that in order to apply the phase correction, $c(\info_N)$, one has to compute $\ul{A}_N$ using the system's parameters such as $\kappa, \gu, \gd, \theta_n, \tau_n$. If there are imperfections in the knowledge of these system parameters, so they can only be estimated with uncertainty, then it is possible that the coherence vector $\ul{A}_N$ used in computing the phase correction and its true value (in the underlining dynamics) are not the same.  In this section, we will give a numerical calculation method to calculate the coherence with imperfections. This method provides a benchmark for the analytical results that we will derive in the following sections for all imperfections.

There are now two sets of parameters: the true values and the estimated values. For our convenience, we will use the previous notation for variables $\{ \kappa,  \Theta, \tau \}$ for the estimated values (or what an experimenter thinks their values should be) and we will use $\{ \kappa + \delta \kappa,  \Theta + \delta \Theta, \tau + \delta \tau \}$ for the true values which are differed from the estimated ones by some unknown values $\delta \kappa, \delta \Theta, \delta \tau$. With this notation, let us define a new coherence vector $\underline{A}_{N}^{\delta}$ to describe the true underlining coherence dynamics of the system. The phase-corrected coherence in Eq.~\eqref{eq:oldcoh_ctrl} becomes:
\begin{align} \label{eq:CtrlCoh_expandwithimp}
    \coh_{c(\bullet)}^{\delta} &=\left|\sum_{Y_N}e^{-ic(Y_N)} \underline{I}^\top\,  \underline{A}_{N}^{\delta} \right|,
\end{align}
where $c(\info_N)={\rm arg}\, \ul{I}\tp \ul{A}_N$ follows the optimal control formula, but with the coherence vector $\ul{A}_N$ computed from the estimated parameters, while $\ul{A}_N^{\delta}$ represents the true evolution of the coherence vector with uncertainties. For example, if there is uncertainty in the measurement angle, $\delthe$, the evolution of $\underline{A}_N^{\delta}$ should be governed by the matrix $\mathbf{F}(\mu_n = \{ s(\Theta_n + \delthe), \tau_n\}, y_n)$ with the uncertainty $\delthe$, while the final phase correction has to be calculated from the matrix $\mathbf{F}(\mu_n = \{ s\Theta_n, \tau_n\}, y_n)$ with an assumed-perfect (but estimated) value of $\Theta$. Therefore, in this case, the control can not be expected to be optimal and the expected coherence cannot be maximized. For all the system imperfections we will consider in this paper, the numerical results of the expected coherence can all be calculated using \eqref{eq:CtrlCoh_expandwithimp}. The summation in \eqref{eq:CtrlCoh_expandwithimp} means that we sum all the possible readout trajectories $\info_N$ ($2^{N}$ strings of readout), and take the average of the coherence for each trajectory by their corresponding probabilities. We therefore denote this method as sum over all paths (SOP) hereafter.

\section{Uncertainty in measurement angles $\delthe$}\label{sec:delthe}

In this section, we consider a scenario where, even though an experimenter aims to set $\theta_{n+1} = s_n \Theta$ for the adaptive algorithm as in Eq.~\eqref{eq:adtmoa}, there can be uncertainty in the actual measurement angle, which turns out to be
\begin{align}
    \theta_{n+1} = s_n (\Theta + \delthe),
\end{align}
with an error $\delthe$ that is assumed unknown to the experimenter. Since the SQ's measurement results depend on the actual angle implemented, the likelihood function for the measurement result in Eq.~\eqref{eq:forwardP} should be modified as:
\begin{align} \label{eq:probdelthe0}
      \wp^{\delthe}(y_n|&\theta_n, x_n)=y_n + (-1)^{y_n}\cos^2\sq{\frac{1}{2}(s_n(\Theta+\delthe)-\ks x_n)},
\end{align}
where $s_n = {\rm sgn}(\zeta_n)$ is chosen adaptively as before. Given this modified likelihood function, any non-zero error $\delthe \ne 0$ will break the condition Eq.~\eqref{eq-probone} in the perfect case. That is, whenever the algorithm guess the RTP correctly $s_n = z_n$, there will still be a non-zero probability that we have a NNR to occur,  \begin{align}\label{eq-probonefalse}
   \wp^{\delthe}(y_{n+1} = 1| \theta_{n+1}, x_{n+1}) =
   \sin^2(\delthe/2) \ne 0,
\end{align} 
by substituting $\theta_{n+1} = s_n (\Theta + \delthe)$ and $x_{n+1} = z_n \Theta/K$ in Eq.~\eqref{eq:probdelthe0}. These false NNRs are wrong indicators that the RTP $z(t)$ were flipped, thereby contributing to an increase in the decoherence of the DQ. 

In order to calculate the decoherence as a result of this measurement angle uncertainty, we first need to analyze all possible cases of the false NNRs and how they affect the coherence of the DQ. Let us assume that the uncertainty is reasonably small compared to the MOAAAR optimal measurement angle, \ie, $|\delthe| \ll \Theta^*$. Then we can expect that false NNRs usually appear in pairs. The reason for that is as follows. If, at a time $t_n$, there occurs one false NNR $y_n = 1$ from Eq.~\eqref{eq-probonefalse}, this will consequently flips $s_{n} = - z_n$. This $s_n$ is then used for the next measurement at $t_{n+1}$, where the measurement result $y_{n+1}$ occurs with probabilities:
\begin{align}\label{eq:probdelthe}
    \wp^{\delthe}(y_{n+1} &= 0 | \theta_{n+1}, x_{n+1}) = \cos^2\sq{ z_n(\Theta+\delthe/2)}, \\
    \wp^{\delthe}(y_{n+1} &= 1 | \theta_{n+1}, x_{n+1}) =   \sin^2\sq{ z_n(\Theta+\delthe/2)}.
\end{align}
One can see that, with the chosen angle $\Theta = \Theta^* \approx 1.5$ and $\delthe \ll \Theta^*$, the probabilities above give $\wp^{\delthe}(y_{n+1}=1|\theta_{n+1}, x_{n+1}) \approx 1$, which means that it is highly likely to obtain the second NNR at $t_{n+1}$. If this NNR actually occurs, it will flip back $s_{n+1} = z_{n+1}$ to agree with the RTP and the situation is back to what was before the first false NNR. This way, we can think of the second NNR as a correction of the first. In Fig. \ref{fig:tBandw}, the bottom panel (blue dots) illustrates the situation of the false pair of NNRs, where we only show the case with $z(t)=+1$. 

It is also important to note that, in between the pair of NNRs mentioned above, there can be arbitrary numbers of null results occurred. This is expected from Eq.~\eqref{eq:probdelthe}. Instead of getting an immediate second NNR, there is a small chance to get a null result $y_{n+1} =0$. Since the null result does not change the value of $s_n$, the probabilities in Eq.~\eqref{eq:probdelthe} will still be applied to the following measurements at $t_{n+2}$, $t_{n+3}$, ..., until the second NNR finally occurs. Therefore, the number of null results between the pair of NNRs can be anything from 0 to $\infty$. Note that we only consider cases in the asymptotic regime and the time period we are interested is infinite long.


Assuming that $\delthe$ is small and so the probability of a false NNR, we can consider all events that only include a single pair of NNRs with arbitrary numbers of null results in between (higher numbers of pairs are unlikely). 
Let us denote $T_2 = N_2 \tau$ as a time duration of interest that include only a single pair of NNRs and define the time between the pair as $w\tau$, where $w-1$ is the number of null results between the pair and $N_2$ goes to $\infty$.  Examples of strings of $N_2 = 10$ measurement results for $w = 1,2,3$ are
\begin{align}\label{eq:n210examples}
Y_{w=1}=&\{1,1,0,0,0,0,0,0,0,0\},\nonumber \\
Y_{w=2}=&\{1,0,1,0,0,0,0,0,0,0\},\nonumber \\
Y_{w=3}=&\{1,0,0,1,0,0,0,0,0,0\}.
\end{align}
noting that we chose the first false NNR (the first `1') to locate at the beginning of the arrays for illustrative purpose, while in general it can locate anywhere in time. The example shown in Fig. \ref{fig:tBandw} was a case of $w=2$, where the duration between the pair of false NNRs is $2\ddt$. 

To analyze all possible cases of the false NNRs, we then calculate the probability of $w$, which can be obtained from the probability of the strings $\wp(Y_w)$ using the modified likelihood function \eqref{eq:probdelthe0}. The probability of $w$ given that the NNR pair occurs during the time $T_2$ is 
\beq\label{eq:Probw}
\begin{aligned}
    \wp(w| &\, 2 {\rm NNRs} \in T_2)=\frac{1}{\wp(2 {\rm NNRs} \in T_2)}\sum_{Y_w}\wp(Y_w),
\end{aligned}
\eeq
where the summation is over all possible combinations of measurement results $Y_w$ for a particular $w$. To compute $\wp(Y_w)$, we use the fact that the probability of null results before and after the NNR pair is $\cos^2(\delthe/2)$, which is almost 1 and thus 
$\wp(Y_w)$ for $w>0$ can be approximated by a product of probabilities of: (a) the first false NNR, (b) the $w-1$ null results in between, and (c) the last correcting NNR. That is, for $w>0$, 
\begin{align}\label{eq-probyw}
    \wp(Y_w) &= \wp^{\delthe}(y_{(a)}=1|\theta_{(a)},x_{(a)})\times \left[\wp^{\delthe}(y_{(b)}=0|\theta_{(b)},x_{(b)})\right]^{w-1}\nonumber\\
    &\times \wp^{\delthe}\left(y_{(c)}=1|\theta_{(c)},x_{(c)}\right),\nonumber\\
    &= \sin^2(\delthe/2)\left[\cos(\Theta+\delthe/2)\right]^{2(w-1)}\sin^2(\Theta+\delthe/2) 
\end{align}
where we have used
\begin{align}
    \theta_{(a)} &= -\theta_{(b)} =  -\theta_{(c)} = z_n(\Theta + \delthe),\\
    x_{(a)} &= x_{(b)} = x_{(c)} = z_n \Theta/K,
\end{align}
which are results of the adaptive algorithm. To evaluate the summation over $Y_w$ in Eq.~\eqref{eq:Probw}, we see that the contribution from $w \ne 1$ in Eq.~\eqref{eq-probyw} is insignificant when $\delthe$ is small. For example, the probability $\wp(Y_w)$ for $w > 2$ is approximately $10^{-6}$ for $\delthe=0.2$. This means that the contribution from $Y_{w=1}$ in Eq.~\eqref{eq:n210examples} dominates the rest of $Y_{w>1}$.
If we only consider the dominant contributions and that the $\wp(Y_w)$ in Eq.~\eqref{eq-probyw} does not explicitly depend on the location of the first NNR,  then the summation over all possible combinations of $Y_w$ will simply be $N_2$ times its individual value because the number of possible $Y_{w=1}$ is $N_2$.
Similarly, the normalized factor $\wp(2 {\rm NNRs} \in T_2)$, which is the probability of having a pair of NNRs in the duration of time $T_2 = N_2\tau$, can be computed as
\begin{align}\label{eq-normprobw}
    \wp(2 {\rm NNRs} \in T_2) \approx N_2 \sin^2(\delthe/2),
\end{align}
from $N_2$ multiplied by the probability of only the first NNR.
Therefore, we can now obtain Eq.~\eqref{eq:Probw} and calculate $\langle w^2\rangle$ as:
\begin{equation}\label{eq:meanofwsquare}
    \langle w^2\rangle=\frac{N_2}{2}\sin^2(\delthe/2)\left( 3+\cos(2\Theta+\delthe) \right)\csc^4(\Theta+\frac{\delthe}{2}),
\end{equation}
which will be useful in computing the coherence of the DQ in this case.

To find the coherence of DQ at an arbitrary time $T_1$ as defined in Eq.~\eqref{eq-beftimecoh}, we need to first compute the mean square error between the real phase, $\phid(T_1)=\kd T_1-2\kd t_{\rm B}$, as in Eq.~\eqref{eq:reDQphase}, and the estimated phase $\varphi_{T_1}$ for this case with the uncertainty in the measurement angle. In this calculation, we can restrict to the case where $z(t)$ starts from $+1$ and remains for a long time then jumps to $-1$ without loss of generality since we consider a symmetric case $\gu=\gd$ in this work. Together with the effect from the Before-time in Eq.~\eqref{eq:esDQphase}, the estimated phase in an arbitrary $T$ (or we can simply consider this is the total time including $t_{\rm B}$ and $w$ events as this time will not change the decoherence rate) with the false pair of NNRs can be calculated as: 
\beq\label{eq:esDQphasew}
\varphi_{T}=\kd T-2\kd \langle t_{\rm B} \rangle-2w\kd\ddt,
\eeq
where the last term, $-2w\kd\ddt$, is from the (possibly occurring) pair of false NNRs. Because for a pair of NNRs, there are $w$ intervals during which the estimated noise $s_n$ becomes $-1$ instead of the real value $+1$. Therefore the difference is always $2$ for each interval, and the total estimated phase should be calculated together with the corresponding slopes, $\lambda_{s_{n+1}}^{y_{n+1}}$. The exact calculation of $\lambda_{s_{n+1}}^{y_{n+1}}$ is given in \ref{sec:appendix1}, and we approximate it as $1$ to obtain the last term, $-2w\kd\ddt$ in \eqref{eq:esDQphasew}. (In the case where $z(t)$ starts from $-1$, the term would be $+2w\kd\ddt$, but this gives the same results since we only care about the mean of $w^2$). 
Since the estimated phase is now dependent on $w$, the mean square error should be calculated by averaging, not only over the distribution of $t_{\rm B}$, but also over the distribution of $w$. This leads to
\beq\label{eq:MSEphasedth}
\begin{aligned}
 \left\langle\left[\phid(T)-\varphi_{T}\right]^2\right\rangle_{t_{\rm B},w}\!\!=4\kd^2\left(\langle t_{\rm B}^2\rangle-\langle t_{\rm B}\rangle^2 \right)+4\kd^2\ddt^2\langle w^2\rangle,
\end{aligned}
\eeq
using the expectation values found in Eq.~\eqref{eq:varoftB} and Eq.~\eqref{eq:meanofwsquare}. Note that the mean of $w$ does not appear because the MOAAAR algorithm does not take into account that a pair of false NNRs may have occurred. A different algorithm that did take that into account could therefore perform better. But it would require the size of $\langle(\delta\Theta)^2\rangle$ (consider a Taylor expansion of Eq.~(\ref{eq-probyw}), and we are not assuming that knowledge).  

Using the similar technique as in Eq.~\eqref{eq:decoresult}, we then approximate the final decoherence rate as 
\beq
\begin{aligned}
\bar{\Gamma}^{\delthe}(\Theta)&\approx \frac{4\kd^2(\langle t_{\rm B}^2\rangle_{t_{\rm B}}-\langle t_{\rm B}\rangle_{t_{\rm B}}^2)\bg T_1}{2T_1}+\frac{4\kd^2 \langle w^2\rangle_w \ddt^2}{2T_2} \\
&\approx\frac{\kd^2}{2\ks^2}\left\{H_\Theta+\frac{\ks}{\bg}\frac{\left(\delthe\right)^2}{2}\Theta\left[ 3+\cos(2\Theta) \right]\csc^4(\Theta)\right\}, \label{eq:decoratedth}
\end{aligned}
\eeq
keeping terms up to the second order in $\delthe$. In the above equation, we have used an approximation sign even in the first line because of two main reasons. First is that we have only considered effects from one pair of false NNRs, since $\delthe$ is small compared to $\Theta$. From \eqref{eq-normprobw}, we see that the probability to have more than one pair of false NNRs is in the order of $\mathcal{O}(\delthe^4)$, which we neglected. The second reason is that we have approximated the Before-time $t_{\rm B}$ as independent of the false NNR pair, \ie, not related to the variable $w$. This is not true in general because, if $\delthe$ is not small, there might be more than one pair of false NNRs which can overlap in time with the real jump of RTP, \ie, the time $t_{\rm B}$ would be affected by the false NNRs. 

\begin{figure}
    \centering
    \includegraphics[width=0.7\textwidth]{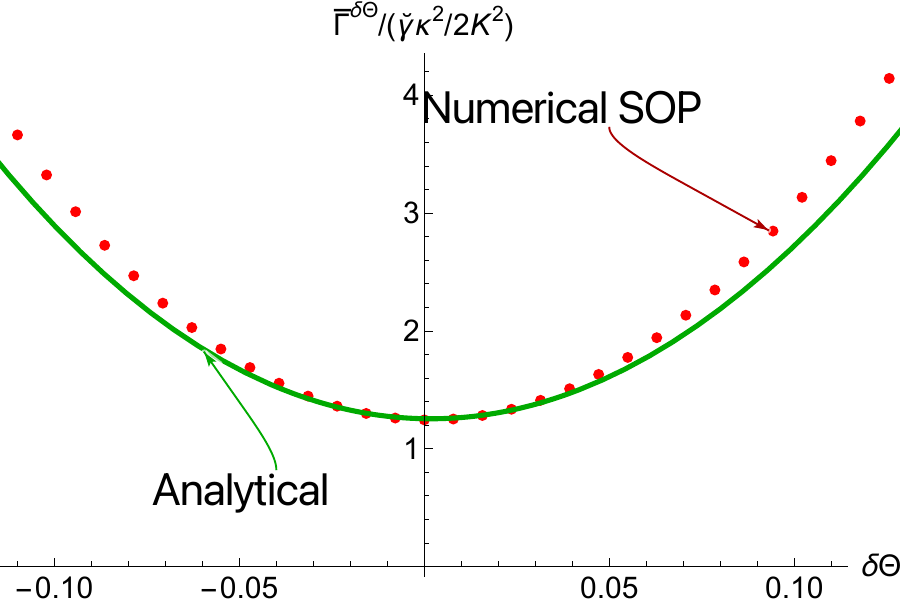}
    \caption{The scaled decoherence rate of DQ are plotted as a function of measurement angle uncertainty, showing both analytical results (Eq.~\eqref{eq:decoratedth}, solid green) and numerical results (red points). The numerical results are generated from summing over all possible paths (SOP) using the mapping matrices $F_s^y$ with $\Theta+\delthe$. To obtain the numerical data points, we generated all the possible $\info$ strings according to their possibilities to compute the average decoherence as a function of time. We then extracted the decoherence rate in the asymptotic regime from the slopes decoherence versus time, throwing away the first five points to avoid transient effects. The parameters are $\kd=0.2$, $K=100$, $\gu=\gd=1$, and $\Theta = \Theta\opt \approx 1.5$.}
    \label{fig:scaledGammadth}
\end{figure}

Next, we analyze the scaling of the decoherence rate by looking at the relative size of the two terms in the first line of Eq.~\eqref{eq:decoratedth}. 
The contribution from $\delthe \ne 0$ in the second term is amplified by the additional factor $K/\bg$, meaning it could dominate the value of $H_\Theta$ even when $\delthe$ is small. By taking $H_\Theta$ and any $\Theta$-dependent functions in those terms to be finite and independent of other parameters, we can conclude that, in order for both terms in Eq.~\eqref{eq:decoratedth} to have the same scaling, we need $\ks(\delthe)^2/\bg= \mathcal{O}(1)$, \ie, the angle uncertainty should satisfy $\delthe = \mathcal{O}(\sqrt{\bg/\ks})$.

We compare our analytical prediction in Eq.~\eqref{eq:decoratedth} with the numerical results that are generated from the SOP method presented in Sec.~\ref{sec:sop}, where the map matrices $F_s^y$ now include the measurement angle uncertainty. As shown in Fig.~\ref{fig:scaledGammadth}, the prediction  matches quite well with the numerical result for small $\delthe$ and starts to diverge for large values of $\delthe$. This is as expected since in the analytical calculation, we have made an assumption which requires small $\delthe$, as discussed above.  

\section{Uncertainty in DQ's noise sensitivity $\delta \kappa$}\label{sec:delkappa}
In this section, we consider the effect of uncertainty in the DQ's noise sensitivity denoted by $\delkd$. Similar to the previous section, this uncertainty could occur when an experimenter mistakenly believe that the sensitivity was $\kappa$, where the actual value was $\kappa +\delkd$. We define the coherence of phase-corrected data qubit following Eq.~\eqref{eq:oldcoh_ctrl}, where we have replaced $\kappa$ with $\kappa + \delkd$ to get
\begin{subequations}
\begin{align}
\coh_{c(\info)}^{\delkd} &:=\,\, \left|\left\langle e^{i [(\kd+\delkd) \xd - c(\info)]}\right\rangle_{\xd,\info} \right|, \label{eq:cohdelkda}\\
&\approx \,\, 1-\frac{1}{2}\langle[\kd \xd-c(\info)]^2\rangle_{\xd,\info}-\frac{(\delkd)^2}{2}\langle\xd^2\rangle_\xd -\delkd\langle \xd[\kd \xd-c(\info)]\rangle_{\xd,\info}\label{eq:cohdelkdb},    
\end{align} 
\end{subequations}
and, in the second line, we have expanded the exponential term to the second order in $\delkd$.
We can evaluate all terms in Eq.~\eqref{eq:cohdelkdb} from the results we have so far. Let us start with the second term in \eqref{eq:cohdelkdb}, which is the exponential part of the control coherence for the ideal case, $\delkd = 0$.
Thus the average decoherence rate in the asymptotic time for the adaptive protocol with the measurement angle $\Theta$ (see Ref.~\cite{song2023optimized}) is given by 
\begin{align}\label{eq:cpdelkd1}
\bar{\Gamma}(\Theta)
\approx \frac{1}{2 T}\langle[\kd \xd-c(\info)]^2\rangle_{\xd,\info},
\end{align}
where $T$ is the final time when the coherence is evaluated. Similarly, for the third term in Eq.~\eqref{eq:cohdelkdb}, we have the no-control average decoherence rate defined as
\begin{align}
    \Gamma^{\rm nc} \approx \frac{1}{2 T}\langle \kd^2\xd^2\rangle_\xd.
\end{align}
Finally, the last term in \eqref{eq:cohdelkdb} can be obtained by applying the result from the perfect case. Assuming the regime of interest where Eq.~\eqref{eq-beftimecoh} is still applied, we can derive that $\langle \kd \xd\rangle_{\xd|\info} \approx c(\info)$ and $\langle\kd \xd-c(\info)\rangle_{\xd,\info} \approx 0$ for the optimal control $c(\info)$. This gives,
\begin{align}
\left\langle \xd[\kd \xd-c(\info)]\right\rangle_{\xd,\info}& \approx \frac{1}{\kd}\left[ \langle \kd^2\xd^2\rangle_{\xd,\info}-\left\langle \langle \kd\xd \rangle_{\xd|\info} c(\info)\right\rangle_{\info}\right]\nonumber\\
&=\frac{1}{\kd}\left\langle [\kd \xd- c(\info)]^2\right\rangle_{\xd,\info}
\end{align}
which is exactly the same term as the one in Eq.~\eqref{eq:cpdelkd1}. Therefore, by combining all the results above into \eqref{eq:cohdelkdb}, we obtain the average decoherence rate for the case with uncertainty in the DQ's noise sensitivity, $\delkd$, as:
\beq\label{eq:decohratedelkd}
\bar{\Gamma}^{\delkd}(\Theta)\approx \bar{\Gamma}(\Theta)+2\frac{\delkd}{\kd}\bar{\Gamma}(\Theta)+\frac{\delkd^2}{\kd^2}\Gamma^{\rm nc},
\eeq
where the last two terms are contributions from the $\delkd$ uncertainty. The second term in Eq.~\eqref{eq:decohratedelkd} exhibits a linear relationship, implying that the uncertainty directly translates to a proportional change in the decoherence rate. The last term in \eqref{eq:decohratedelkd}, however, is of greater significance and tends to dominate the linear term in most relevant regimes due to the presence of $\Gamma^{\rm nc}$. Using the expression of $\Gamma^{\rm nc}$ in Eq.~\eqref{eq:gammanc}, we can obtain the uncertainty bound for $\delkd$ at which the decoherence rate scales with $K$ as in the perfect case. Specifically, we require that
\begin{equation}\label{eq:kappabound}
\frac{\delkd^2}{\kd^2} \Gamma^{\rm nc}=\frac{\delkd^2}{\kd^2}\frac{\kd^2\bg}{2\bar{\gamma}^2} \sim \frac{\kd^2\bg}{2\ks^2}H(\Theta).
\end{equation}
Since $H(\Theta) = \mathcal{O}(1)$,  
we obtain $\delkd/\kd = \mathcal{O} (\bar{\gamma}/\ks)$ as the upper bound of the uncertainty $\delkd$.

\begin{figure}
    \centering
    \includegraphics[width=0.7\textwidth]{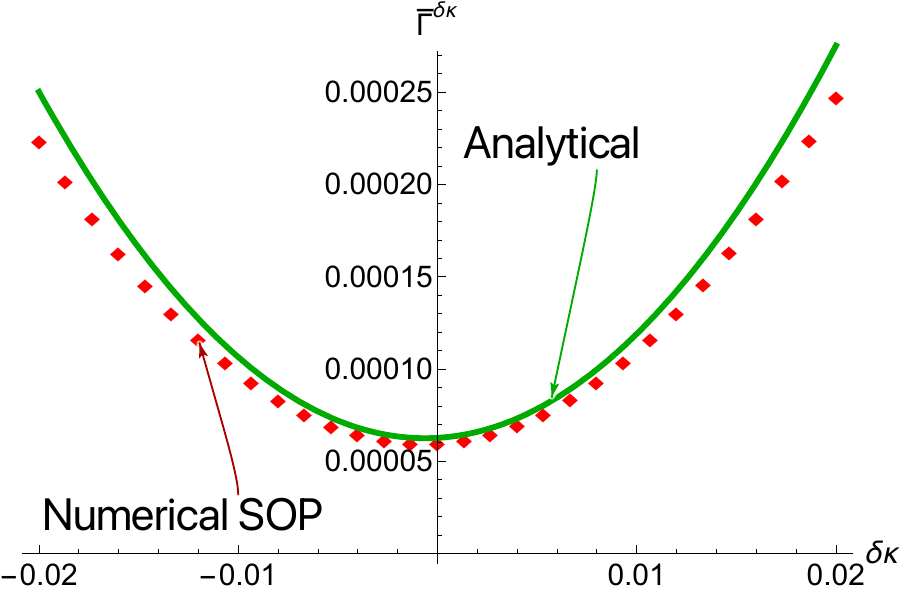}
    \caption{The decoherence rate of DQ is plotted as a function of the uncertainty $\delkd$, showing both analytical and numerical results. The analytical result (solid green) is from \eqref{eq:decohratedelkd} and the numerical result (red points) was generated from summing over all possible paths (SOP), using the mapping matrices $F_s^y$ with $\kd+\delkd$. The decoherence rate is extracted in the similar way as in Fig.~\ref{fig:scaledGammadth}. The parameters we used are: $\kd=0.2$, $K=20$, $\gu=\gd=1$, and $\Theta = \Theta\opt \approx 1.5$.}
    \label{fig:decodelkd}
\end{figure}

We show in Fig.~\ref{fig:decodelkd} a comparison between the predicted decoherence rate $\bar{\Gamma}^{\delkd}(\Theta\opt)$ in \eqref{eq:decohratedelkd} and the numerical results from the SOP method for $\Theta = \Theta\opt \approx 1.50055$. In this figure, we use $K=20$ instead of $K=100$ as in other plots. This is because, with a finite value of uncertainty $\delkd$, it takes more time for the decoherence to reach its linear regime (\ie, the decoherence grows linearly in time, away from the transient regime) and so choosing a smaller $K$ with a longer measurement time $\ddt=\Theta/K$ can help reduce the total number of measurements required to reach the long time limit. As expected, the numerical results match reasonably well with our analytical prediction for small $\delkd$, with a slight deviation (even at $\delkd=0$) from the fact that $K=20$ is not large enough to be in the asymptotic regime.

\section{Imperfection relevant to measurement times}\label{sec:meastime}
In this section, we consider imperfections that are related to the measurement times of the SQ. It was assumed in the previous work that, after each of the probing times $\ddt$, the SQ could be measured (where a result $y_n$ was taken) and reset instantaneously, as well as any detection process related to the SQ can be called in operation at any point in time (infinite bandwidth). We here relax these assumptions and assume that the SQ requires additional time during the readout and reset processes (spectator readout and reset time) and that there is a dead time during which the SQ cannot be measured because the detection process is not ready.

\begin{figure}
    \centering
    \includegraphics[width=0.7\textwidth]{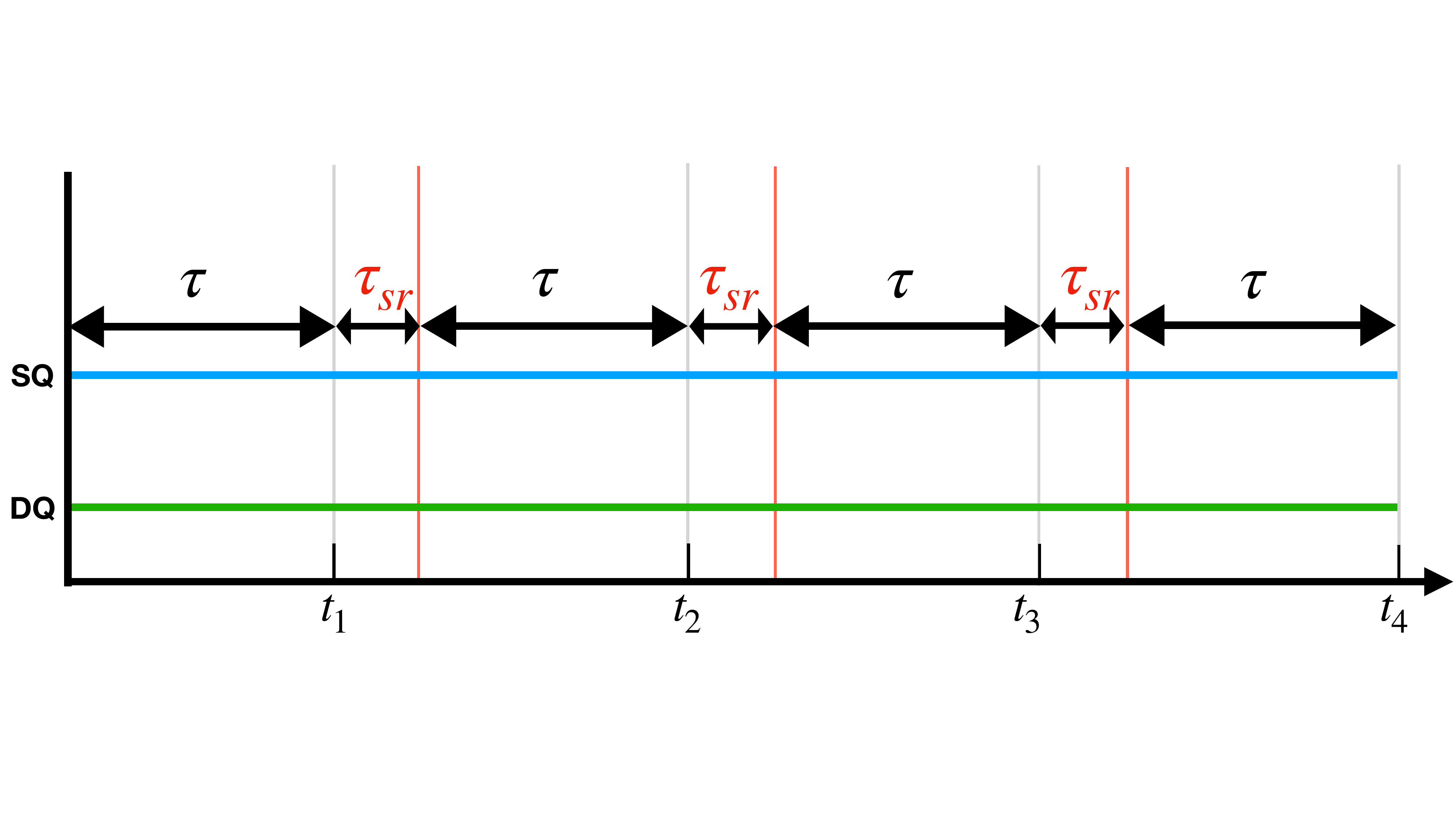}
    \caption{Schematic diagrams showing the SQ's probing ($\tau$) and reset ($\tsr$) times. After the SQ probes the noise for time $\tau$ and is measured at $t_n \in \{ t_1, t_2, t_3, ...\}$, the SQ will be reset to its initial state. However, if the reset process is not instantaneous and takes some time, $\tsr$, then the SQ will not be functioning to probe the RTP noise during that time, causing the degradation in decoherence of the DQ.}
    \label{fig:sqreset}
\end{figure}

\subsection{SQ's readout and reset time}\label{sec:sqreset}
Let us denote the SQ's readout and reset time as $\tsr$, as the time during which the SQ is being read out and reset to its initial state. During this time, the SQ should be assumed as not functioning at all as a noise sensor. We illustrate this imperfect time in Fig.~\ref{fig:sqreset}, where the non-zero $\tsr$ will delay the noise probing process after every measurement. Since the waiting/probing time in the adaptive algorithm is in the order of $\ddt= \mathcal{O}(1/\ks)$, it is reasonable to assume that the reset time is of the same order. Therefore, we can define
\begin{align}
\tsr=\frac{\coeftsr}{K},
\end{align}
where we have denoted $\coeftsr\sim\mathcal{O}(1)$ as a constant factor determining the size of $\tsr$.

It is straightforward to compute the coherence and the decoherence rate for this case, using the generalized map-based formalism as introduced in Section~\ref{sec-genmap}. Since the SQ does not sense noise during this reset time, but the DQ still evolves under the RTP, a coherence vector $\ul{A}_n$ at time $t_n$ for this case should be updated using a combination of ${\bf H}(\tsr,\kd)$ and ${\bf F}(\mu_n = \{ \theta_n,\ddt_n\},y_n)$. The former describes the evolution of coherence without measurement (the SQ is not sensing the noise) and the latter describes the evolution during which the SQ is sensing the noise, with a readout $y_n$. If we still attempt to execute the MOAAAR adaptive strategy with $\theta_{n+1}=s_n\Theta$ and $\ddt_{n+1} = \tau \equiv \Theta/K$, then after $N$ measurements, the coherence vector will be given by 
\begin{align}\label{eq:covectsr_update}
    \underline{A}_N= \left[\prod_{j=1}^N {\bf{H}}(\tsr,\kd){\bf{F}}(s_j\Theta,\ddt,y_j) \right]\underline{A}_0,
\end{align}
using the definitions of both matrices as in Eqs.~\eqref{eq:Hmatrix} and \eqref{eq:Fmatrix}. With the coherence vector, one can compute DQ's coherence as well as its decoherence rate, \eg, numerically using the SOP method.

\begin{figure}
    \centering
    \includegraphics[width=0.7\textwidth]{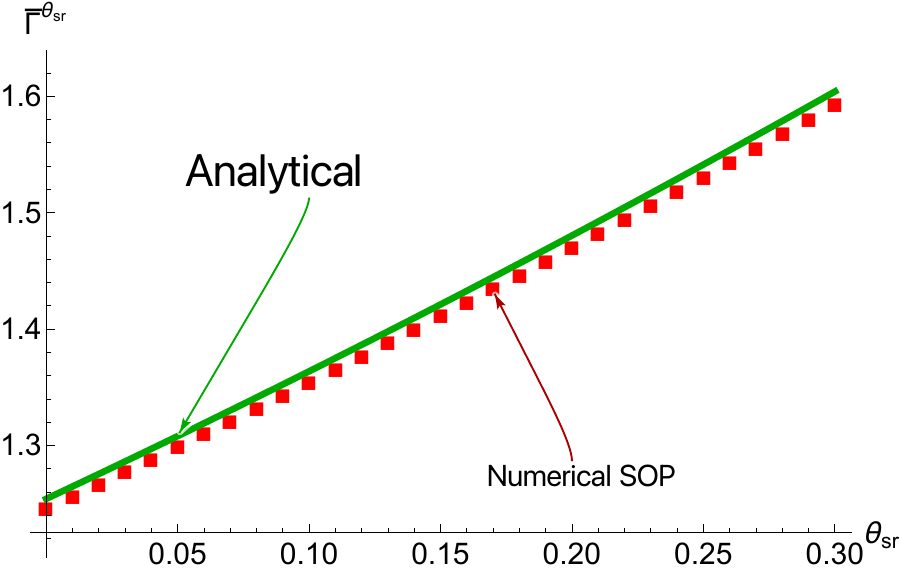}
    \caption{Plots of decoherence rate as a function of $\coeftsr$. The solid green line shows the analytical result in \eqref{eq:decoratetsr}, and the red square points are degenerated by SOP. We use $\kd=0.2, \ks=100,\gu=\gd=1, \Theta\approx 1.50055$.}
    \label{fig:decotsr}
\end{figure}

Fortunately, in this case, we can obtain a closed-form expression for the decoherence rate in the asymptotic regime. We follow the procedure proposed in Ref.~\cite{tonekaboni2023greedy}, looking for stable eigenstates of the dynamics where the values of $\zeta_n$ and $\alpha_n$ defined in Eqs.~\eqref{defzeta} and \eqref{defalpha} converge to only particular values after about $N=10$ measurements. 
We found that, for any value of $\coeftsr$,  the dynamics exhibit two most occupied states, which correspond to the stable eigenstates of the combined mapping matrix defined as 
\begin{align}
\newF_{s}^y(\Theta,\coeftsr) \equiv \hfunc(\tsr = \coeftsr/K,\kappa)\ffun(s\Theta,\ddt,y).
\end{align}
Let us denote the eigenstates of the combined matrix $\newF_{s}^y(\Theta,\coeftsr)$ by $\ul{E}_s^0(\coeftsr)$ for $s\in\{-1,+1\}$. By using these stable eigenstates, we are able to calculate the expected decoherence rate as (see Ref.~\cite{tonekaboni2023greedy} for more details):
\begin{equation} \label{eq:asymav}
\begin{aligned}
\bar\Gamma^{\theta_{\rm sr}}(\Theta) := \sum_{s=\pm} {P}_{\rm ss}(s)\frac{1}{\ddt | \underline{I} ^\top \underline{E}_s^0 (\coeftsr)|}\left[| \underline{I}^\top \underline{E}_s^0(\coeftsr) | - \sum_{y} | \underline{I} ^\top\, \newF_{s}^y(\Theta,\coeftsr) \, \underline{E}_s^0(\coeftsr) | \right].
\end{aligned}
\end{equation}
By solving for $\ul{E}_{\,+}^{0}(\coeftsr)$ and $\ul{E}_{\,-}^{0}(\coeftsr)$ and keeping terms up to $\mathcal{O}(1/K^2)$, we obtain 
\begin{align}\label{eq:decoratetsr}
\bar{\Gamma}^{\theta_{\rm sr}}(\Theta)=H_\Theta^{\theta_{\rm sr}}\frac{\bg\kd^2}{2\ks^2},
\end{align}
where we have defined a new prefactor,
\beq\label{eq:Htsr}
\begin{aligned}
H_\Theta^{\theta_{\rm sr}}&=H_\Theta+{\coeftsr}\left[\frac{2\Theta}{3}-8\Theta\csc^2\Theta+8\Theta\csc^4\Theta \right]+{(\coeftsr)}^2\left[ \frac{1}{3}-4\csc^2\Theta+4\csc^4\Theta \right],    
\end{aligned}
\eeq
with $H_\Theta$ from the perfect case in Eq.~\eqref{eq:Hfunction}.
For the decoherence rate in Eq.~\eqref{eq:decoratetsr} to have the same scaling as that of the perfect case, we thus need $\coeftsr = {\cal O}(1)$, which means that $\tsr = {\cal O}(1/\ks)$. Again, we can compare the analytical results with the numerical ones obtained from the SOP method, where the agreement can be seen in Fig.~\ref{fig:decotsr}.

\subsection{Dead time in SQ's detection process}\label{sec:deadtime}
\begin{figure}
    \centering
    \includegraphics[width=0.7\textwidth]{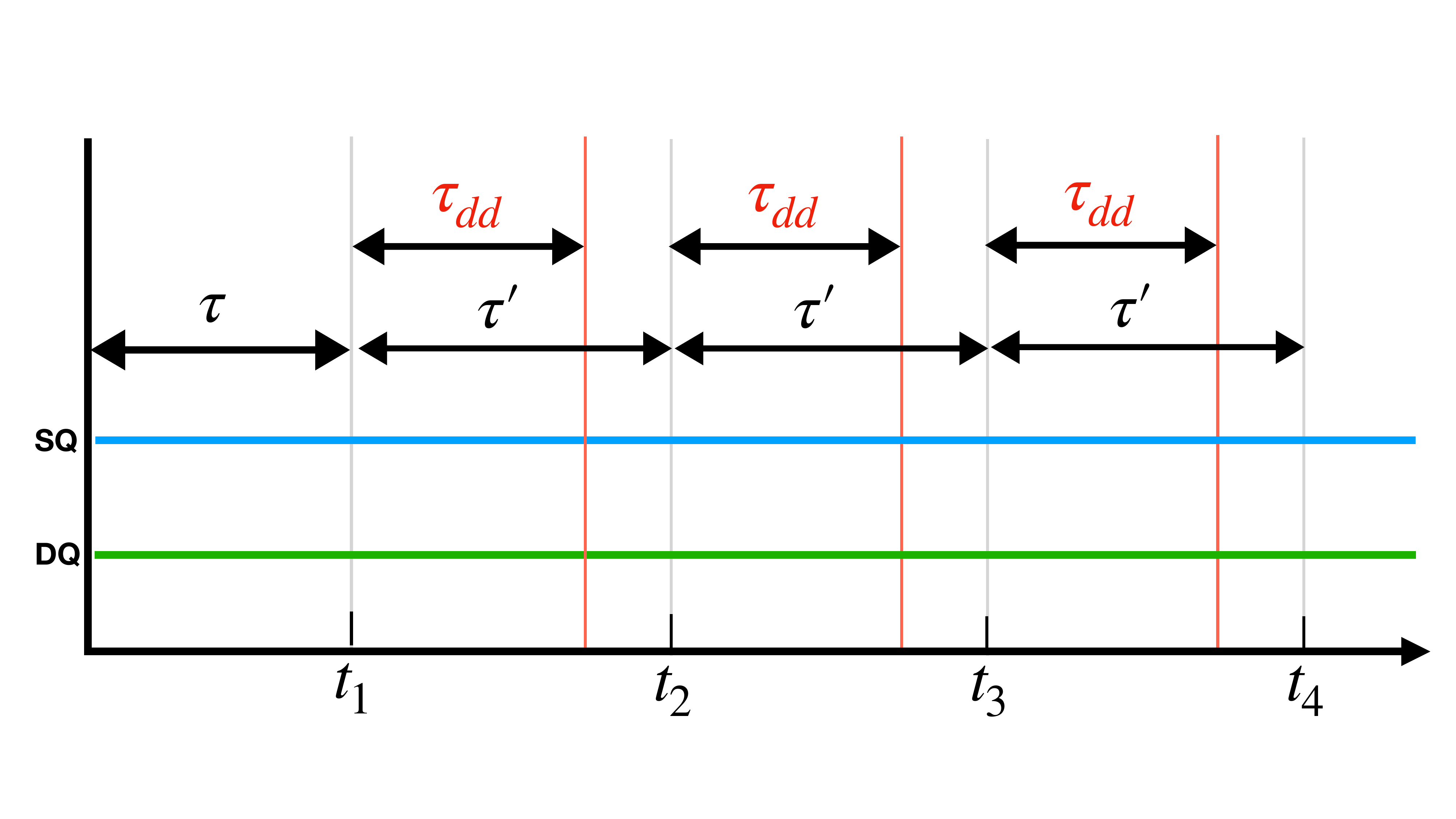}
    \caption{Systematic depict of detector's dead time. The optimal waiting time between two measurements is $\ddt$. However, when the detector is not ready to measure the SQ in $\ddt$ time and can only be available in $\deadt>\tau$ after each measurement, the best time to conduct next measurement is not immediately after $\deadt$ but $\tau^\prime$.}
    \label{fig:ddtime}
\end{figure}

Another time imperfection is when any detection process of the SQ is not functioning, causing a dead time where the SQ cannot be measured. In this case, even though the best waiting time for the adaptive algorithm was known to be $\ddt=\Theta\opt/\ks \approx 1.50055/\ks$, the SQ's detection process might not be ready yet and the measurement time has to be delayed further. We refer to Fig.~\ref{fig:ddtime} for the illustration that the dead time leads to a delay in measuring the SQ. Let us denote this dead time as $\deadt$. If the dead time is shorter than the optimal time, \ie, $\deadt<\ddt$, then this imperfection will not affect our results as the measurement time can still be chosen as $\ddt$. However, when $\deadt\geq\ddt$, the measurement time has to be delayed to a new waiting time denoted by $\ddt^\prime$. To calculate the coherence of the data qubit, we update the coherence vector using only the matrix, $\ffun(s\Theta,\ddt^\prime,y)$, but with the new waiting time $\ddt^\prime$ and still choosing $\Theta=\Theta\opt = \ks \tau$. Here we keep $\Theta$ fixed in order to study the effect of just one alteration to MOAAAR, but below we consider changing $\Theta$ to match $K\tau'$. We will see below when the dead time is beyond a specific value, it is better to wait for a longer time that satisfies $\ddt^\prime \approx (\Theta\opt + n \pi)/K \geq \deadt$ rather than measuring immediately at $\tau' =\deadt$.

\begin{figure}
    \centering
    \includegraphics[width=0.75\textwidth]{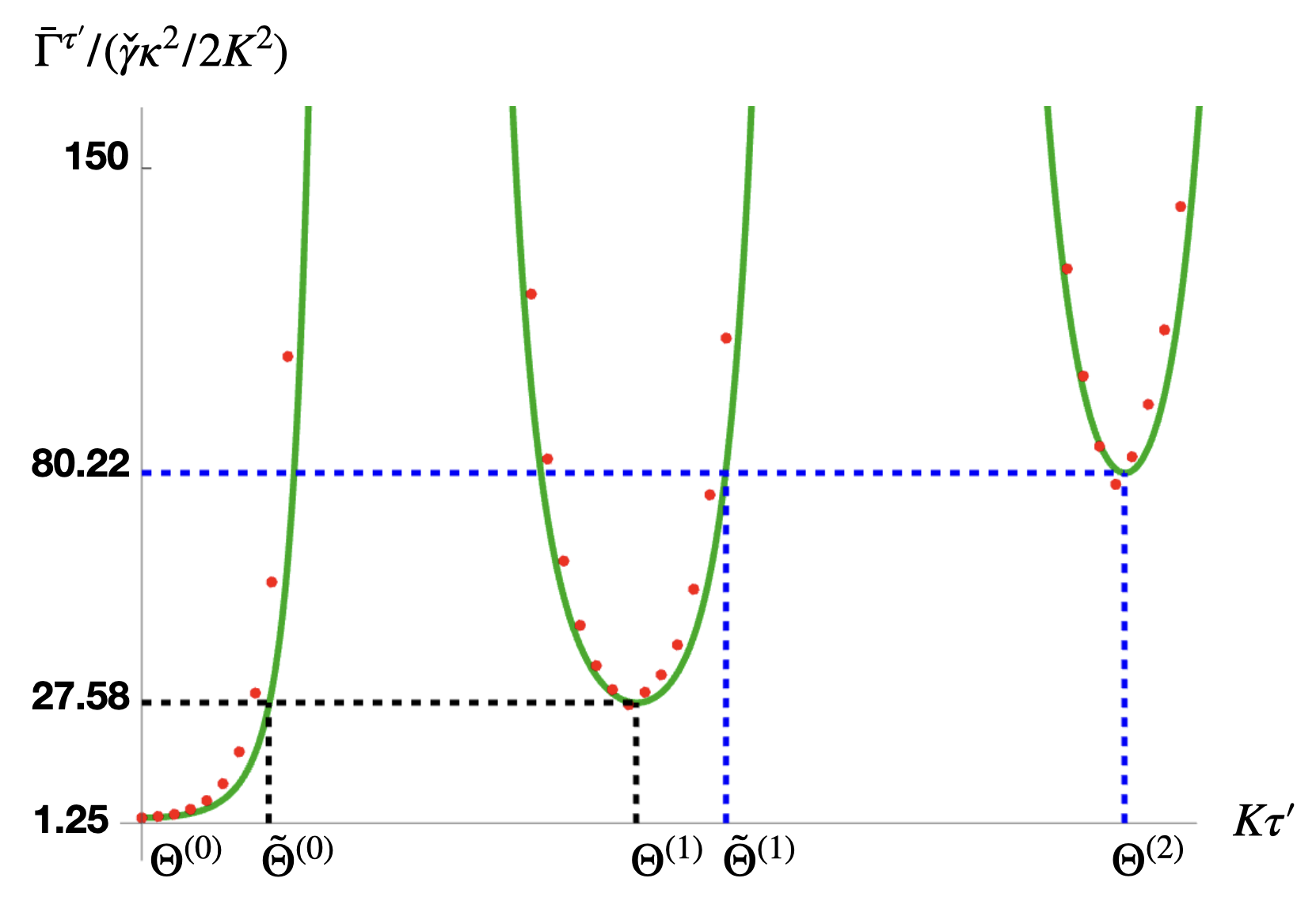}
    \caption{Scaled decoherence rate with different waiting time $\ddt^\prime$ for both analytical and numerical results. We use the parameters $\kd=0.2,\ks=100,\gu=\gd=1$. The numerical results are shown in red points which are generated using SOP, and the analytical result is simply plotting the ideal $H_\Theta$ for $\Theta=\ks\ddt^\prime$. We do not expect these two results to match perfectly since they are representing different situations, and the detailed explanation can be found in text. The black dashed line shows the time of $\ks\ddt^\prime=\tilde{\Theta}^{(0)} \approx 2.325$, by which we should measure the SQ immediately otherwise we should choose to wait for longer time until approximately $\ks\ddt^\prime=\Theta^{(1)}\approx 4.69$ to get the smaller decoherence rate.}
    \label{fig:decorateddt}
\end{figure}

In order to find a proper waiting time to measure the SQ to minimize the decoherence rate of the DQ, the SOP method is required to generate numerical values, as analytical predictions are not available in this case. With the map $\ffun(s\Theta,\ddt^\prime,y)$, we use the SOP method to find the numerically averaged decoherence over time and then extract its slopes in time to obtain the decoherence rate, denoted by ${\bar \Gamma}^{\rm dd}$. We generate the decoherence rate for different values of the waiting time in the range of $\tau' \in [\Theta\opt/K, 3\pi/K]$, using the parameters: $\kd=0.2$, $\kd=100$, $\gu=\gd=1$, $\Theta=\Theta\opt \approx 1.50055$. We show (as red data points) in Fig.~\ref{fig:decorateddt} the numerical decoherence rate scaled by the factor $2 \ks^2/(\bg \kd^2)$. Even though we do not have a theoretical prediction, we can still compare our numerical results with the scaled decoherence rate for the perfect (no imperfection) adaptive case described by $H_\Theta$ in Eq.~\eqref{eq:Hfunction}.  We plot the scaled rate $H_\Theta$ (corresponding to $\Theta$ matching $\ks\tau^\prime$) as the green solid curve in Fig.~\ref{fig:decorateddt}.   The agreement between the numerical values and the function $H_\Theta$ is quite surprising, even though one has to realize that the former assumed the fixed measurement angle $\Theta = \Theta\opt$ independent of the waiting time $\tau'$, while the latter allowed the angle to change with the waiting time according to the adaptive algorithm relation $\Theta = K\tau'$.

The agreement between the numerical decoherence rate and the function $H_\Theta$ in Fig.~\ref{fig:decorateddt} suggests that we can use $H_\Theta$ as a guide to approximate an appropriate waiting time to measure the SQ in the presence of the dead time. It is clear from the plot that $\tau' = \tau \equiv \Theta\opt/K$ is the best option to minimize the decoherence rate (shown as the graph's origin in Fig.~\ref{fig:decorateddt}, where $\Theta^{(0)} = \Theta\opt$). However, if the dead time is longer than the optimal time, \ie, $\deadt > \tau$, then the optimality has to be compromised and the SQ can only be measured when the detection process is ready at $\tau' \ge \deadt$, resulting in the decoherence rate increases. Until the dead time exceed a threshold value at $\tau' = \tilde\Theta^{(0)}/K = 2.32/K$ (see the black dashed lines in Fig.~\ref{fig:decorateddt}), instead of measuring the SQ when ready, it is better to wait for a longer time to implement the measurement at  $\ddt^\prime\approx \Theta^{(1)}/\ks =  4.69/\ks$. It it interesting that this similar strategy repeats when the dead time gets even longer, as shown by the blue dashed lines in Fig.~\ref{fig:decorateddt}. That is, when the dead time exceeds a threshold at  $\ddt^\prime \approx \tilde\Theta^{(1)}/\ks = 5.27/\ks$, then it is better to wait and measure at $\ddt^\prime \approx \Theta^{(2)}/\ks = 7.84/\ks$. We can therefore summarize the strategy as
\begin{align}\label{eq:deadtstrategy}
    \tau' \approx \begin{cases} \Theta^{(i+1)}/\ks, & \text{if}\quad \ks\deadt \in [\tilde\Theta^{(i)}, \Theta^{(i+1)}] \\ \deadt, &  \text{otherwise},\end{cases}
\end{align}
for $i = 0, 1, 2, ...$. For convenience, we have defined $\Theta^{(i)}$ as an $i$th root of the derivative function $\partial_\Theta H_\Theta$, \ie,
\begin{align}
    \Theta^{(i)} \in \{ \Theta \,:\, \partial_\Theta H_\Theta = 0\},
\end{align}
where the first root is the optimal MOAAAR angle, $\Theta^{(0)} = \Theta\opt$. We have also defined $\tilde\Theta^{(i)}$ as another measurement angle that gives the same function value as $H_{\Theta}$ for $\Theta = \Theta^{(i+1)}$. 
Therefore, we need to know a rough idea of the range of the detector dead time $\deadt$ in order to implement the measurement accordingly to reduce the decoherence rate.

To see the effect of the detector's dead time on the decoherence rate, we can look at the value of $H_\Theta$. For example, if the detector is not ready by $\tau' = \tilde{\Theta}^{(0)}/\ks$, then it is better to wait until $\tau' = \Theta^{(1)}/\ks$. It is interesting to note that the latter option is roughly $\Theta^{(1)}/\ks \approx (\Theta^\star+\pi)/\ks$, \ie, the measurement time is delayed by about $\pi/K$ from the optimal value in the ideal case. For the decoherence rate, the delay will increase the value of $H_\Theta$, from $H_{\Theta^\star}\approx 1.254$ to $H_{\Theta^{(1)}}\approx H_{\Theta^{\star}+\pi}\approx 27.58$, \ie, almost 22 times larger. Similarly, if the detector's dead time is even longer, $\tau' > \tilde{\Theta}^{(1)}/\ks$, one should choose $\tau' = \Theta^{(2)}/\ks \approx (\Theta^\star+2\pi)/\ks$, which is roughly $2\pi/K$ delayed from the optimal value. This will increase the value of $H_\Theta$ to $H_{\Theta^{(2)}}\approx H_{\Theta^{\star}+2\pi} \approx 80.22$. We note that, although the decoherence rates may seem worse compared to the perfect case, these are the best options available when the detector's dead time becomes a major issue in the experiments.

\section{Measurement error and additional SQ dephasing}\label{sec:measerr}
In this section, we move on to the last imperfection, which is the SQ's measurement inefficiency and the additional dephasing on the SQ. We will show that these two affect the coherence of the DQ similarly. 

We first consider the measurement inefficiency, where there is an  error probability of $\measerr$ that an outcome $y = 1$ mistakenly appeares as $y = 0$ or vise versa. We use the observable given in Eq.~\eqref{eq:thetaop} to define the effect operator $\hat {E}_y$ for the perfect case, where $\hat {E}_0 \equiv \hat{\Pi}_\theta$ and $\hat E_1 \equiv {\mathbb{1}}-\hat{\Pi}_\theta$ are for null and non-null results, respectively, using $\hat{\Pi}_\theta=|\theta\rangle^{\rm s}\langle\theta|$ as a projector. Therefore, we can modify the operators to include the measurement error to obtain
\beq\label{eq:measoperators}
\begin{aligned}
\hat{E}_0^{\measerr} \equiv(1-\measerr)\,\hat{\Pi}_\theta+\measerr\,(\mathbb{1}-\hat{\Pi}_\theta),\\    
\hat{E}_1^{\measerr} \equiv (1-\measerr)\, (\mathbb{1}-\hat{\Pi}_\theta)+\measerr\, \hat{\Pi}_\theta.
\end{aligned}
\eeq
Using these effect operators, we obtain the probability function for a measurement result $y$ from
\begin{align}\label{eq:probymeaserr}
\wp^{\measerr}(y|\theta,x) = &\, ^{\rm s}\langle \ks x|\, \hat{E}_y^\epsilon\,|\ks x\rangle^{\rm s} ,\nonumber \\
= &\,y+(-1)^{y}\cos^2\left[\frac{1}{2}(\theta-\ks x)\right]-(-1)^{y}\measerr\cos\left(\theta-\ks x\right),    
\end{align}
which will reduce to the perfect case in Eq.~\eqref{eq:forwardP} for $\measerr = 0$. The above probability function indicates that, even if we can choose an optimal adaptive measurement angle, $\theta=s\Theta$, such that it matches perfectly with the SQ's real phase, \ie, $s\Theta=z\ks \ddt$ (assuming that the adaptive algorithm has guessed $s = z$ correctly), there is still a probability $\measerr$ to get a NNR, $y=1$.

\begin{figure}
    \centering
    \includegraphics[width=0.9\textwidth]{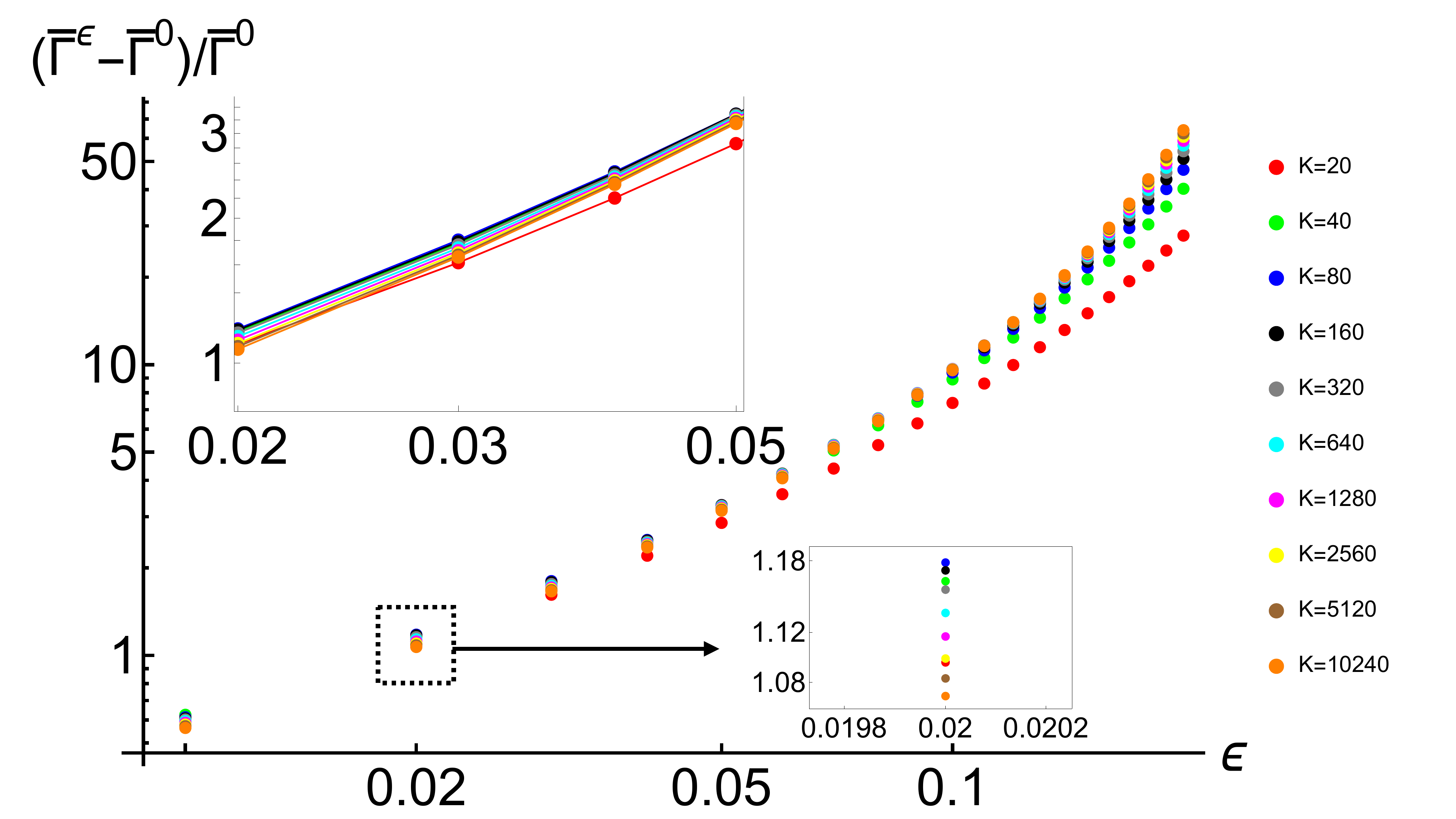}\caption{The decoherence rates change with measurement error rates $\epsilon$. In the vertical axis, $\bar{\Gamma}^0$ is the decoherence rate with $\measerr=0$, which also indicates the result of the perfect case. To obtain the decoherence rate, we first use the SOP in Section~\ref{sec:sop} to calculate the expected decoherence with time. Then the decoherence rate is extracted from the slopes by throwing away the first $10$ measurement steps (of $18$ measurement steps in total) to avoid the transient effects. The top-left inset is a zoomed-in plot, for $\epsilon\leq 0.04$, showing an approximately power-law (linear on this log-log plot) relation between  the difference ratio $(\bar{\Gamma}^\measerr-\bar{\Gamma}^0)/\bar{\Gamma}^0$ and  $\epsilon$. Another inset in the right bottom shows the convergence of the decoherence rate with the increase of $\ks$, except for $\ks=20$ and $\ks=40$ that are not in the asymptotic regime. We keep fixed the parameters $\gu=\gd=1$ and $\kd=0.2$ for different $\ks$ and $\measerr$.}
    \label{fig:miplot}
\end{figure}

We then consider the additional dephasing of the SQ, which can be modeled as an exponential decay of the off-diagonal elements of the SQ's density matrix.  Because we reset the SQ to its initial state $|\Phi=0\rangle^{\rm s}$ after every measurement, the additional dephasing will only affect it during the waiting time $\ddt$. Thus, the SQ's state just before the next measurement becomes
\beq
\rho_{\rm s}(\sqdepha)=\frac{1}{2}
\begin{bmatrix}
1& e^{-\sqdepha \ddt}e^{-i\phis}\\
e^{-\sqdepha \ddt}e^{i\phis}&1
\end{bmatrix},
\eeq
where the off-diagonal elements are multiplied by the dephasing factor $e^{-\sqdepha \ddt}$, with a denoted dephasing rate $\sqdepha$. Assuming a perfect projective measurement $(\measerr = 0)$, we obtain the probability function for a measurement result $y$ as:
\begin{align}\label{eq:probysqdephase}
\wp^{\sqdepha}(y|\theta,z)=&\,  ^{\rm s}\langle \theta + \pi y |\rho_{\rm s}(\sqdepha)|\theta + \pi y\rangle^{\rm s}\nonumber \\
= & \, y+(-1)^{y}\cos^2\left[\frac{1}{2}(\theta-\ks x)\right]-(-1)^{y}\frac{\sqdepha\ddt}{2}\cos\left(\theta-\ks x\right),   
\end{align}
where we have only kept terms up to ${\cal O}(\sqdepha)$ assuming that $\sqdepha$ is small. By comparing the probability functions in Eq.~\eqref{eq:probysqdephase} and Eq.~\eqref{eq:probymeaserr}, we find that they are the same when $\measerr=\sqdepha\ddt/2$. Moreover, if there are both imperfections occurred at the same time, then the total effect will be described by the same probability function replacing $\measerr$ by $\measerr + \chi \ddt/2$. In the following, without loss of generality, we will only analyze $\measerr$ and its effects on the DQ's decoherence.

To compute the decoherence for this case, we modify the mapping matrix based on the new likelihood function in Eq.~\eqref{eq:probymeaserr} and get
\beq\label{eq:Fwithmeaserror}
\begin{aligned}
\newFepsilon^{\measerr}\left(\theta,\ddt,y\right):=& \frac{1}{4}\Big[2 \, \hfunc(\ddt,\kd)  +(-1)^{y}(1-2\measerr) \sum_{\xi=\pm1} e^{-i\xi\theta}\hfunc(\ddt,\kd+\xi\ks) 
\Big]    
\end{aligned}
\eeq
where the detailed calculation can be found in~\ref{sec:appendix2}.
We then apply the SOP method using the modified matrix $\newFepsilon^{\measerr}(\theta,\ddt,y)$ to obtain  numerical results of the DQ's coherence. Let us define $\bar{\Gamma}^0$ and $\bar{\Gamma}^\measerr$ as the average decoherence rate for the perfect case $\measerr=0$ and the imperfect case $\measerr\ne 0$, respectively. We show numerical results in Fig.~\ref{fig:miplot}, where we plot the difference ratio $(\bar{\Gamma}^\measerr-\bar{\Gamma}^0)/\bar{\Gamma}^0$ on a log-log scale as a function of the error $\measerr$. Even though it does not seem to have any simple relationship, we can notice that, for a small error, $\measerr<0.05$, the curve is approximately linear and independent of the sensitivity $\ks$. With this observation, we can extract the linear relation:
\begin{align}
    \log\left(\frac{\bar{\Gamma}^\measerr-\bar{\Gamma}^0}{\bar{\Gamma}^0}\right)=a\log(\measerr)+b,
\end{align}
and obtain $a\approx 1$ and $b\approx 4$, which means that
\begin{equation}
\begin{aligned}
\bar{\Gamma}^\measerr&=\bar{\Gamma}^0\left(1+e^{4}\measerr\right) \approx \bar{\Gamma}^0\left(1+50\measerr\right). \label{eq:decoratedepsilon}
\end{aligned}
\end{equation}
Therefore, if we require that the decoherence rate for the imperfect case to be double that of the perfect one, then we need $1+50\epsilon \approx 2$, which is satisfied when $\measerr \lesssim 0.02$. The insets in Fig.~\ref{fig:miplot} show zoom-in plots where we can see the data points for different values of the sensitivity $\ks$.

\begin{figure}[t]
    \centering
    \includegraphics[width=0.49\textwidth]{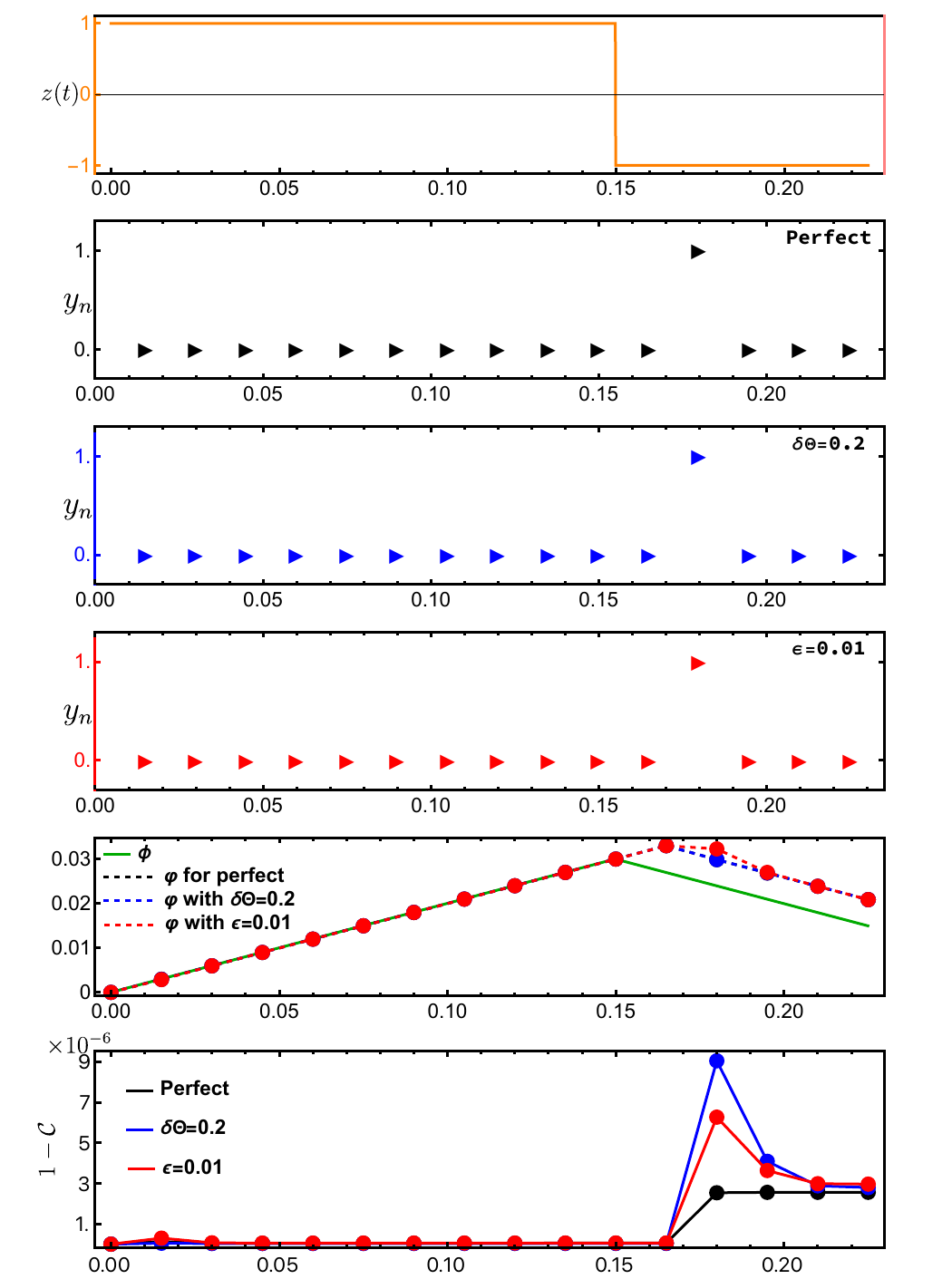}
    \includegraphics[width=0.49\textwidth]{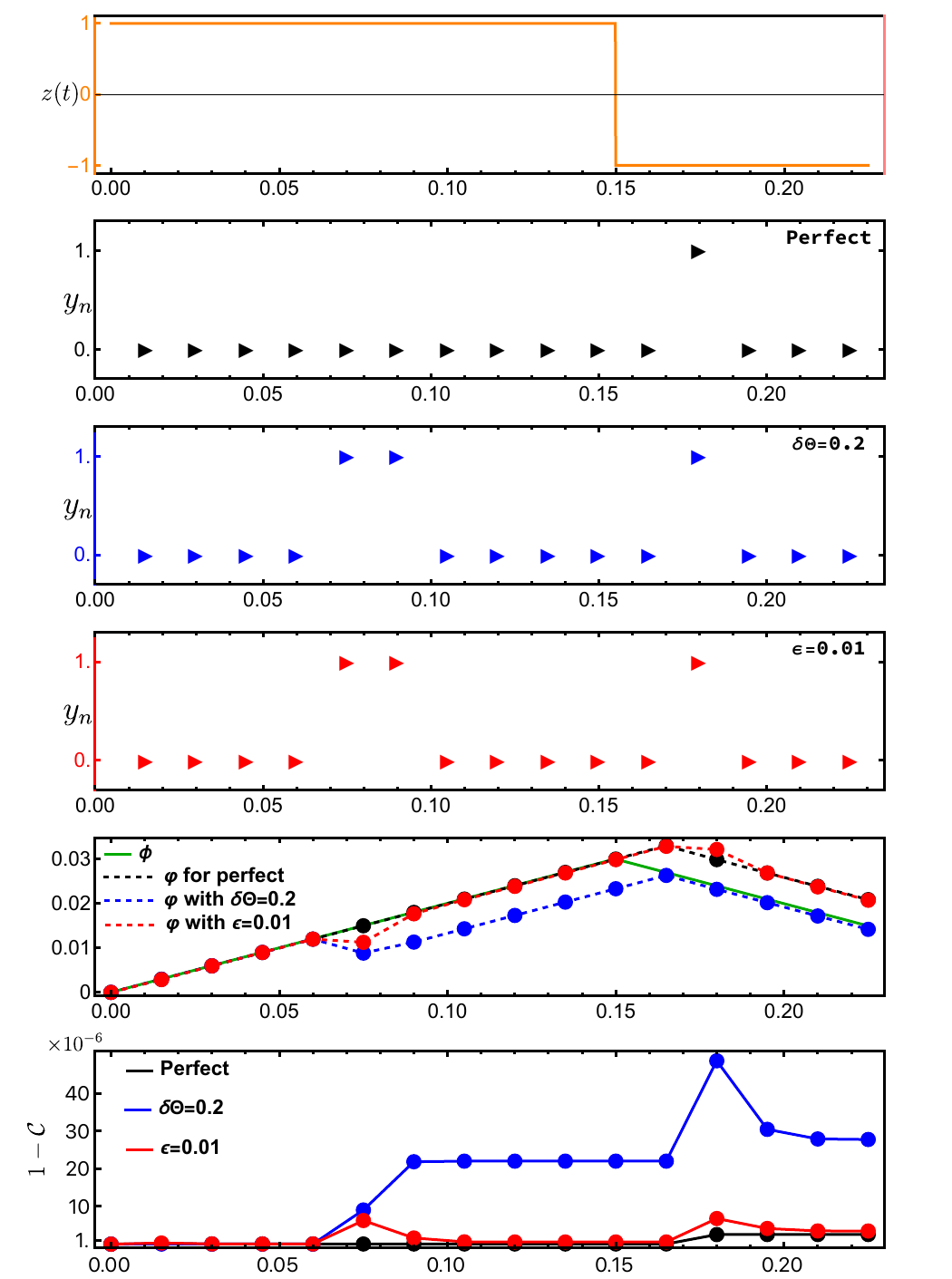}
    \caption{The two possible cases for a fixed RTP noise. In both left and right figures, we show the readout strings of $y_n$ for the perfect case, the cases with measurement angle uncertainty $\delthe$ and measurement error $\measerr$, and their corresponding estimated phases and decoherence with time. The parameters we use are $\kd=0.2, \ks=100,\gd=\gu=1, \Theta=\Theta\opt \approx 1.50055, \ddt=\Theta\opt/\ks,\measerr=0.01$ and $\delthe=0.2$, which gives $\measerr\approx\sin^2(\delthe/2)$. In the left column, we show that even though the readout strings for the three cases are the same, their estimated phase and decoherence are different. While in the right column, we show that there is possibility of getting the wrong NNRs for both $\delthe$ (blue triangles) and $\measerr$ (red triangles). We note that the wrong NNRs do occur in pairs for both cases, where the latter NNR is a result of correcting the first false NNR as discussed in Sections~\ref{sec:delthe} and \ref{sec:measerr}. The decoherence calculated in the last row is the expected (Bayesian) value, based on knowledge of the measurement outcomes and the dynamical parameters (including the true values of $\measerr$ and $\delthe$). That is, it is not the actual decoherence in this particular run, as would be calculated from the true DQ phase, $\phi$ (green), resulting from the RTP, in the second-last row.}
    \label{fig:individualplot}
\end{figure}

It is also interesting to note the measurement error $\measerr$ in this section and the uncertainty in the measurement angle in Section~\ref{sec:delthe} both have similar effects on the probability of measurement results as described in Eqs.~\eqref{eq:probymeaserr} and \eqref{eq:probdelthe0}, respectively. By comparing the two probability functions, we recognise that they coincide when the probability to obtain the NNR $y=1$ is given by
\begin{align}\label{eq-comparetheerror}
    \sin^2(\delthe/2)=\measerr.
\end{align}
That is, the probability of the NNR is equal to the measurement error $\measerr$. Despite this, these two types of imperfections lead to different values of the decoherence, as can be seen by comparing Eq.~\eqref{eq:decoratedth} to Eq.~\eqref{eq:decoratedepsilon}. To understand how this difference arises, we show some hypothetical individual runs in  Fig.~\ref{fig:individualplot}, plotting the RTP trajectory, the measurement readout strings, the corresponding DQ's phases, and decoherence. We show three cases: with no imperfections; with the angle uncertainty $\delthe = 0.2$; and with the measurement error $\measerr=0.01$, approximately satisfying Eq.~\eqref{eq-comparetheerror}.  

In the left column of Fig.~\ref{fig:individualplot}, we choose the same readout string $Y_N$ for all cases, so no NNRs are obtained during the time when the RTP, $z(t)$, remains unchanged at $+1$. We can see that the corresponding estimated phases $\varphi$ and the decoherence $1-\coh$ in all cases are almost the same. Only when the RTP jumps do we see a clear difference in estimated phase (between the third case and the other two) and decoherence (different for all three). By the end of time series, however, the estimates and decoherence in all three cases have reconverged. In the right column of Fig.~\ref{fig:individualplot}, we show examples of different measurement readout strings, with the same pair of false NNRs in the cases of nonzero $\delthe$ and $\measerr$.
This time, the contributions to the decoherence in these two cases are quite different. The effect of uncertainty $\delthe$ on the final decoherence is far worse than that of the measurement error $\measerr$. This is because the angle uncertainty $\delthe$ is unknown to the protocol while the measurement error $\measerr$ is a known parameter. By looking at the estimated phase and the decoherence in conjunction, 
we can gain insight into how this plays out. 
At the first false NNR, since the $\measerr$ is known and the protocol is aware that the NNR could be a false one, the estimated phase is not diverging from the real phase as far as the case with $\delthe$. (This is, of course, also seen with the genuine NNR in the left column.)  Then at the second false NNR, the protocol with $\measerr$ `realizes' that the first NNR was a false one and immediately returns the phase estimate almost to what it would have been with no false NNRs (black). By contrast, the protocol with $\delthe$ mistakenly takes the second NNR as an indication of a second flip in the RTP. Because of this, the resulting expected decoherence jumps up at both NNRs and remains there, while that for the case $\measerr \neq 0$ jumps up, then down, and then reduces almost to the same level as the no-error case (black).


\section{Conclusion}\label{sec:conclusion}
Building on the results of RTP noise mitigation using SQs under ideal conditions in Refs.~\cite{tonekaboni2023greedy,song2023optimized}, in this work, we have extended the analyses to evaluate the proposed protocol under non-ideal conditions. These included the uncertainty in measurement angles $\delthe$, the uncertainty in the DQ’s noise sensitivity $\delkd$, the SQ’s readout and reset time $\tsr$, the detector dead time $\deadt$, the measurement error $\epsilon$ and additional dephasing $\chi$.
Our findings indicate that when these imperfections remain within certain bounds, the proposed algorithm continues to perform effectively. The main results for these imperfections are discussed below and summarized in Table \ref{table1}. 
\begin{table}[t]
    {\renewcommand{\arraystretch}{1.5}
    \renewcommand{\tabcolsep}{0.85cm}
    \begin{tabular}{|c|c|c|c|c|}
        \hline
        \text{Imperfection}& $\delthe$ & $\delkd/\kd$ & $\tsr$ 
 & $\measerr$ \\
        \hline
        \text{Bound}& $\mathcal{O} \left( \sqrt{\frac{\bg}{\ks}} \right)$ & $\mathcal{O}\left(\frac{\bar{\gamma}}{\ks}\right)$ & ${\cal O}(\frac{1}{\ks})$ & $\lesssim 0.02$\\
        \hline
    \end{tabular}}
    \caption{Summary of imperfections and their bounds analyzed in this work. The imperfections include uncertainty in measurement angles ($\delthe$), uncertainty in the DQ's noise sensitivity ($\delkd$), a SQ reset time ($\tsr$), and errors in SQ's readout ($\epsilon$) equivalent to added SQ dephasing. The bound is defined as the order of imperfection that allows the decoherence to remain approximately the same as that achieved in the perfect case (see details in the main text). The bound $\measerr\lesssim0.02$ was obtained from the numerical simulations, with $\gu=\gd$, for when the imperfect decoherence rate be double that of the perfect case.}
    \label{table1}
\end{table}

The uncertainty in the measurement angle, $\delthe$, 
induces false NNRs in the SQ's measurement. These lead our algorithm, MOAAAR, to conclude wrongly that the RTP flipped, thereby contributing to an
increase in the decoherence of the DQ after correction. 
Nevertheless, if $\delthe$ is bounded to the order of $\sqrt{\bg/\ks}$, the decoherence rate can still be mitigated to roughly the same degree as in the perfect case. As for the uncertainty in the DQ's noise sensitivity, $\delkd$, it directly increases the DQ's decoherence rate as described in Eq.~\eqref{eq:decohratedelkd}, which is independent of the measurement and control protocol. We found that if the ratio between the sensitivity uncertainty and the real sensitivity is in the order of $\bar{\gamma}/\ks$, the MOAAAR protocol can still suppress the DQ's decoherence in the same scaling as the perfect case. 

For the SQ's readout and reset time, $\tsr$, which is a duration at which the noise information cannot be probed by the SQ, its effect on the DQ's decoherence rate is equivalent to that of the DQ's pure dephasing, as if the SQ does not exist at all. We show analytically that to maintain the same decoherence scaling as in the perfect case, the reset time $\tsr$ must be bounded to $\mathcal{O}(1/\ks)$. For an error $\epsilon$ in the SQ readout, we obtain the bound $\measerr \lesssim 0.02$ from numerical simulations with equal transition rates $\gu=\gd$, but a similar threshold would exist for other jump rates. (For these simulations we chose $\kd/\bar\gamma=0.2$, but the result will not depend on this parameter.) 
Both $\tsr$ and $\measerr$ should be known imperfections. Thus, while we apply the original MOAAAR to the SQ for simplicity, we take into account the values of these two parameters when calculating the final phase correction to the DQ.

For the detector's dead time, $\deadt$, during which the SQ can sense the noise but itself cannot be measured, if the dead time exceeds the optimal time required to measure the SQ, $\deadt > \tau$, the waiting time between measurements increases, thus directly raising the decoherence rate. However, by analyzing the function of scaled decoherence rate, $H(\Theta)$, we found that there was a threshold time, beyond which the measurement time should be delayed until $\Theta^{(i+1)} \approx (\Theta\opt + (i+1)\,\pi)/\ks$, where $i = 0, 1,...$, such that the decoherence rate can still be minimized as summarized in Eq.~\eqref{eq:deadtstrategy}. Finally, we found that extra dephasing of the SQ has a similar impact on the DQ's decoherence rate as measurement error.

Given our analyses, if the imperfections in experiments are within the given bounds, the decoherence suppression should still stay quadratic in the SQ sensitivity as the ideal case. For the future work, one can extend our analyses by replacing the single RTP noise with more general multi-level noises or continuous-value noises (\eg, Gaussian or Ornstein–Uhlenbeck process noise). Finally, working with practical experimental setups may reveal other differences or imperfections that need to be considered.

\section*{Data availability statement}
All data that support the findings of this study are included within the article (and any supplementary files).


\ack
This work was supported by the Australian Government via the Australia-US-MURI grant AUSMURI000002 and by the Australian Research Council via the Centre of Excellence grant CE170100012. A.C.~also acknowledges the support of the Program Management Unit for Human Resources and Institutional Development, Research and Innovation (Thailand) grant B39G680007. We acknowledge the Yuggera people, the traditional owners of the land at Griffith University on which this work was partly undertaken.

\appendix
\section{Slopes of the estimated phase curves}\label{sec:appendix1}
When we measure the SQ, the value of $s_n$ switches once we obtain a NNR. A typical trajectory of $s_n$ can be seen in Fig.\ref{fig:tBandw}, where the middle row shows the perfect situation and the last row shows the case with measurement uncertainty. Therefore, we can see the $s_n$ as our estimation of the RTP $z(t)$. Suppose now we are at the asymptotic regime and just finish the $n$-th measurement, the accumulated phase of the DQ between the $n$-th and the $(n+1)$-th measurement ($\ddt$ time) can be defined as:
\begin{equation}\label{eq:Deltaphasedef}
\Delta\varphi:=s_{n+1}\elam_s^{y_{n+1}}\kappa \ddt,
\end{equation}
where $\elam_s^{y_{n+1}}$ is the coefficient which satisfies $\elam_s^{y_{n+1}}\leq 1$ and the equality holds if the $s_n$ matches to $z(t)$ exactly and $z(t)$ does not flip. We can understand the $\elam_s^{y_{n+1}}$ as the slope in the phase estimation curves in one measurement gap, which can be seen in the fifth row of Fig.\ref{fig:individualplot}. We use $y_{n+1}$ to distinguish the different coefficients with the different measurement results $1$ and $0$. It is known from the previous paper \cite{tonekaboni2023greedy} that the estimated phase of the DQ in asymptotic regime can be calculate as:
\beq\label{eq:estimatephase}
\varphi_n:= \arg \ccc_{|\info_n} =  \arg\left(A_n^{z=+1}+ A_n^{z=-1} \right),
\eeq
where $A_n^{z=+1}$ and $A_n^{z=-1}$ are the first and second elements of the coherence vector $\ul{A}_n$ at the $n-$th measurement. Therefore, the estimated phase increment from the $n$-th to the $(n+1)$-th measurement can be calculated as:
\beq\label{eq:Deltaphase}
\Delta\varphi:=\varphi_{n+1}-\varphi_n=
\arg \frac{\ul{I}^T\ffun_+^{y_{n+1}}\ul{E}_+^0}{\ul{I}^T\ul{E}_+^0}.
\eeq
Here, $y_{n+1}\in \{0,1\}$ are the possible results for the $(n+1)$-th measurement. The $\ul{E}_+^0$ is an eigenstate of the mapping $\ffun_+^{0}$, and it is the steady state that $\ul{A}_n$ will converge to in the asymptotic regime. By comparing the one-step phase increase in \eqref{eq:Deltaphase} to \eqref{eq:Deltaphasedef}, we obtain four slopes, $\elam_+^0, \elam_-^0, \elam_+^1, \elam_-^1$. 

$\elam_+^0$ is calculated for the case that the coherence vector starts from $\ul{E}_+^0$ at the $n$-th measurement and we obtain a null result, $y_{n+1}=0$. In this case, we have $s_{n+1}=1$ since the noise $z(t)$ remains at $1$ during these two measurements. Similarly, $\elam_-^0$ is calculated if we assume the coherence vector starts from another eigenstate, $\ul{E}_-^0$, and a null result is obtain at the $(n+1)$-th measurement, $y_{n+1}=0$. Then $s_{n+1}=z(t)=-1$ during these two measurements. Using the same methods, we can calculate $\elam_+^1$ and $\elam_-^1$ by considering the situation that a NNR is obtained at the $(n+1)$-th measurement, $y_{n+1}=1$, for $z(t)=+1$ and  $z(t)=-1$, separately.
With the help of \emph{Mathematica}, we obtain:
\beq\label{eq:elam0}
\elam_+^0=\elam_-^0\approx 1-\mathcal{O}(\frac{\mg^2}{\ks ^2}),
\eeq
which proves that during the time without NNRs, the difference between the DQ's real phase and the estimated phase is in the order of $\frac{\mg^2}{\ks^2}$. This can be neglected in the asymptotic regime.
When we obtain a NNR $y_{n+1}=1$, the coefficients can be calculated as:
\beq\label{eq:elam1}
\elam_+^1=\elam_-^1 = \frac{\cot \Theta+\Theta \csc\Theta^2}{\Theta}=-1+\frac{2\langle t_{\rm B}\rangle}{\ddt}.
\eeq
Here we found that the mean of this Before-time, $t_{\rm B}$, is related to the slope of the estimated phase through \eqref{eq:elam1}.

\section{Derivation of $\newFepsilon^\measerr(\theta,\ddt,y)$}\label{sec:appendix2}
We rewrite the modified likelihood function under measurement error $\measerr$ in \eqref{eq:probymeaserr} here for convenience.
\begin{equation*}
\begin{aligned}
\wp(y|\theta,x)&=y+(-1)^{y}\cos^2\left[\frac{1}{2}(\theta-\ks x)\right]-(-1)^{y}\measerr\cos\left(\theta-\ks x\right).     
\end{aligned}
\end{equation*}
From the calculation in previous work \cite{tonekaboni2023greedy}, the definition of elements in $\ffun(\theta,\ddt,y)$ is given by:
\beq\label{eq:elementFfunc}
 F_{z_{t^\prime}}^{z_t}\!\left(\theta,\ddt, y\right) :=\!\!
    \int\!\! dx \,\wp^{\measerr}(y|x,\theta)\,\wp(x,z_t|z_{t^\prime})e^{i \kd x},
\eeq
which is the Fourier transformation of $\wp^{\measerr}(y |\theta, x) \wp(x, z_t | z_{t^\prime})$ from variable $x$ to $\kd$. We use the convolution theorem to calculate the Fourier transform of a product of two functions as
\begin{align}\label{eq:twoFourierT}
F_{z_{t^\prime}}^{z_t} =  \int \!\! \dd k \, \mathcal{F}_{\! x \rightarrow k} [ \wp^{\measerr}(y |\theta, x)] {\cal F}_{x \rightarrow (\kd-k)} [ \wp(x, z_t | z_{t^\prime})], \notag
\end{align}
where  $\mathcal{F}_{x\rightarrow k}[f]$ is defined as $\mathcal{F}_{x\rightarrow k}[f]:= \int \dd x\, e^{i k x}\, f$. The second Fourier transform in \eqref{eq:twoFourierT} is calculated as:
\begin{align}
    \mathcal{F}_{x \rightarrow(\kd-k)} [\wp(x,z_t|z_{t^\prime})] &=\!\! \int \!\! \dd x\,  \wp(x,z_t|z_{t^\prime}) e^{i(\kd - k)x} =H_{z_{t^\prime}}^{z_t}(\tau,\kd -k).
\end{align}
The Fourier transform of the new likelihood function is calculated as:
\begin{align}
\mathcal{F}_{x \rightarrow k} [ \wp^{\measerr}(y | x)] = \frac{1}{4} &\big[2 \delta_{k,0} +(-1)^{y} e^{-i \theta} \delta_{k, -\ks}(1-2\measerr) +(-1)^{y} e^{+i \theta} \delta_{k, +\ks} (1-2\measerr)\big].
\end{align}
With the expressions of these two Fourier transform, we obtain
\begin{align}
    F_{z_{t^\prime}}^{z_t} =  \frac{1}{4} &\big[2 H_{z_{t^\prime}}^{z_t}(\tau,\kd) +(-1)^{y} (1-2\measerr) \sum_{\xi=\pm1} e^{-i\xi\theta}H_{z_{t^\prime}}^{z_t}(\tau, \kd+\xi K) \big], 
\end{align}
which means \eqref{eq:Fwithmeaserror}.


\section*{References}
\bibliographystyle{iopart-num}

\begin{thebibliography}{10}
\expandafter\ifx\csname url\endcsname\relax
  \def\url#1{{\tt #1}}\fi
\expandafter\ifx\csname urlprefix\endcsname\relax\def\urlprefix{URL }\fi
\providecommand{\eprint}[2][]{\url{#2}}

\bibitem{gupta2020adaptive}
Gupta R~S, Edmunds C~L, Milne A~R, Hempel C and Biercuk M~J 2020 {\em npj
  Quantum Information\/} {\bf 6} 53

\bibitem{majumder2020real}
Majumder S, Andreta~de Castro L and Brown K~R 2020 {\em npj Quantum
  Information\/} {\bf 6} 19

\bibitem{song2023optimized}
Song H, Chantasri A, Tonekaboni B and Wiseman H~M 2023 {\em Physical Review
  A\/} {\bf 107} L030601

\bibitem{tonekaboni2023greedy}
Tonekaboni B, Chantasri A, Song H, Liu Y and Wiseman H~M 2023 {\em Physical
  Review A\/} {\bf 107} 032401

\bibitem{singh2023mid}
Singh K, Bradley C, Anand S, Ramesh V, White R and Bernien H 2023 {\em
  Science\/} {\bf 380} 1265--1269

\bibitem{youssry2023noise}
Youssry A, Paz-Silva G~A and Ferrie C 2023 {\em New Journal of Physics\/} {\bf
  25} 073004

\bibitem{lingenfelter2023surpassing}
Lingenfelter A and Clerk A~A 2023 {\em npj Quantum Information\/} {\bf 9} 81

\bibitem{ItaTok2003}
Itakura T and Tokura Y 2003 {\em Phys. Rev. B\/} {\bf 67}(19) 195320
  \urlprefix\url{https://link.aps.org/doi/10.1103/PhysRevB.67.195320}

\bibitem{GalAlt2006}
Galperin Y~M, Altshuler B~L, Bergli J and Shantsev D~V 2006 {\em Phys. Rev.
  Lett.\/} {\bf 96}(9) 097009
  \urlprefix\url{https://link.aps.org/doi/10.1103/PhysRevLett.96.097009}

\bibitem{culcer2009dephasing}
Culcer D, Hu X and Das~Sarma S 2009 {\em Applied Physics Letters\/} {\bf 95}

\bibitem{BerGal2009}
Bergli J, Galperin Y~M and Altshuler B~L 2009 {\em New Journal of Physics\/}
  {\bf 11} 025002 \urlprefix\url{https://doi.org/10.1088/1367-2630/11/2/025002}

\bibitem{viola1999dynamical}
Viola L, Knill E and Lloyd S 1999 {\em Physical Review Letters\/} {\bf 82} 2417

\bibitem{viola2003robust}
Viola L and Knill E 2003 {\em Physical Review Letters\/} {\bf 90} 037901

\bibitem{biercuk2011dynamical}
Biercuk M, Doherty A and Uys H 2011 {\em Journal of Physics B: Atomic,
  Molecular and Optical Physics\/} {\bf 44} 154002

\bibitem{ng2011combining}
Ng H~K, Lidar D~A and Preskill J 2011 {\em Physical Review A\/} {\bf 84} 012305

\bibitem{souza2011robust}
Souza A~M, Alvarez G~A and Suter D 2011 {\em Physical Review Letters\/} {\bf
  106} 240501

\bibitem{medford2012scaling}
Medford J, Barthel C, Marcus C, Hanson M, Gossard A {\em et~al.\/} 2012 {\em
  Physical Review Letters\/} {\bf 108} 086802

\bibitem{paz2013optimally}
Paz-Silva G~A and Lidar D 2013 {\em Scientific Reports\/} {\bf 3} 1530

\bibitem{zhang2014protected}
Zhang J, Souza A~M, Brandao F~D and Suter D 2014 {\em Physical Review
  Letters\/} {\bf 112} 050502

\bibitem{Shor1995}
Shor P~W 1995 {\em Phys. Rev. A\/} {\bf 52}(4) R2493--R2496
  \urlprefix\url{https://link.aps.org/doi/10.1103/PhysRevA.52.R2493}

\bibitem{Steane1996}
Steane A~M 1996 {\em Phys. Rev. Lett.\/} {\bf 77}(5) 793--797
  \urlprefix\url{https://link.aps.org/doi/10.1103/PhysRevLett.77.793}

\bibitem{Terhal2015}
Terhal B~M 2015 {\em Rev. Mod. Phys.\/} {\bf 87}(2) 307--346
  \urlprefix\url{https://link.aps.org/doi/10.1103/RevModPhys.87.307}

\bibitem{paz2017multiqubit}
Paz-Silva G~A, Norris L~M and Viola L 2017 {\em Physical Review A\/} {\bf 95}
  022121

\bibitem{von2020two}
von L{\"u}pke U, Beaudoin F, Norris L~M, Sung Y, Winik R, Qiu J~Y, Kjaergaard
  M, Kim D, Yoder J, Gustavsson S {\em et~al.\/} 2020 {\em PRX Quantum\/} {\bf
  1} 010305

\bibitem{chalermpusitarak2021frame}
Chalermpusitarak T, Tonekaboni B, Wang Y, Norris L~M, Viola L and Paz-Silva G~A
  2021 {\em PRX Quantum\/} {\bf 2} 030315

\bibitem{morello2010single}
Morello A, Pla J~J, Zwanenburg F~A, Chan K~W, Tan K~Y, Huebl H,
  M{\"o}tt{\"o}nen M, Nugroho C~D, Yang C, Van~Donkelaar J~A {\em et~al.\/}
  2010 {\em Nature\/} {\bf 467} 687--691

\bibitem{hanson2007spins}
Hanson R, Kouwenhoven L~P, Petta J~R, Tarucha S and Vandersypen L~M 2007 {\em
  Reviews of Modern Physics\/} {\bf 79} 1217

\end{thebibliography}
\providecommand{\newblock}{}

\end{document}